\pdfoutput=1

\documentclass[11pt]{article}

\usepackage[preprint]{acl}%

\usepackage{xcolor}
\definecolor{ForestGreen}{HTML}{009a56}

\usepackage{times}
\usepackage{latexsym}

\usepackage[T1]{fontenc}

\usepackage[utf8]{inputenc}

\usepackage{microtype}

\usepackage{inconsolata}

\usepackage{graphicx}
\usepackage{multirow}
\usepackage{comment}

\usepackage{amsmath}
\usepackage{amssymb}

\usepackage{subcaption}

\usepackage{times}
\usepackage{fancyhdr,graphicx,amsmath,amssymb}
\usepackage[ruled,vlined]{algorithm2e}
\usepackage{algpseudocode}

\usepackage[draft]{todonotes}   

\title{Private Memorization Editing: Turning Memorization into a Defense to Strengthen Data Privacy in Large Language Models}

\author{
\textbf{
Elena Sofia Ruzzetti$^{1}$, 
Giancarlo A. Xompero$^{1,2}$,} \\
\textbf{Davide Venditti$^{1}$, 
Fabio Massimo Zanzotto$^{1,2}$} \\
$^1$Human Centric ART, University of Rome Tor Vergata, Italy \\
$^2$Almawave S.p.A., Rome, Italy
\\
\small{
\href{mailto:elena.sofia.ruzzetti@uniroma2.it}{\color{black} \tt elena.sofia.ruzzetti@uniroma2.it}}
\\
\small{\href{mailto:fabio.massimo.zanzotto@uniroma2.it}{\color{black} \tt fabio.massimo.zanzotto@uniroma2.it}}
}

\begin{document}

\maketitle

\begin{abstract}
Large Language Models (LLMs) memorize, and thus, among huge amounts of uncontrolled data, may memorize Personally Identifiable Information (PII), which should not be stored and, consequently, not leaked.
In this paper, we introduce Private Memorization Editing (PME), an approach for preventing private data leakage that turns an apparent limitation, that is, the LLMs' memorization ability, into a powerful privacy defense strategy.
While attacks against LLMs have been performed exploiting previous knowledge regarding their training data, our approach aims to exploit the same kind of knowledge in order to make a model more robust.
We detect a memorized PII and then mitigate the memorization of PII by editing a model knowledge of its training data. We verify that our procedure does not affect the underlying language model while making it more robust against privacy Training Data Extraction attacks.
We demonstrate that PME can effectively reduce the number of leaked PII in a number of configurations, in some cases even reducing the accuracy of the privacy attacks to zero.
\end{abstract}

\section{Introduction}

Large Language Models (LLMs) can accurately perform many tasks by extracting information 
and distilling capabilities from their training data.
However, as their size increases, training data becomes more difficult to control and may inadvertently include Personally Identifiable Information (PII) from unaware individuals \cite{miranda2025preserving,italiano2024security,yao2024survey}. Hence, emails, phone numbers, and credit cards can be extracted at inference time by executing privacy attacks \cite{carlini2021extracting, carlini2023quantifying, huang-etal-2022-large}.
Moreover, as LLMs grow in size, 
their chance to verbatim memorize training information increases \cite{nasr2023scalable, ranaldi-etal-2024-investigating, kiyomaru-etal-2024-comprehensive-analysis}. 


Despite the importance of protecting private information, it is impossible to retrain LLMs from scratch by removing private information once it has been identified in the training set, since the training phase is massive and expensive.
Therefore, 
methods that can alter the knowledge of an LLM without further training may help to protect users privacy: 
machine unlearning techniques \cite{yao2024largelanguagemodelunlearning, kassem-etal-2023-preserving} have been successfully applied to preserve users privacy.
Among the most data-efficient ones, model editing methods like Private Association Editing (PAE) \cite{venditti2024enhancingdataprivacylarge} can be targeted to protect a private piece of information. 
In particular, PAE addresses the protection of multiple users with a single edit, breaking the \textit{association} between a user name and its private information.

Interestingly, 
the success of privacy attacks based on verbatim memorized prompts suggests that \textit{LLMs tend to memorize PII rather than associate it with individuals' identity}. 
Indeed, Training Data Extraction attacks \cite{carlini2021extracting,carlini2023quantifying,huang-etal-2022-large,nasr2023scalable} or attacks based on other measures of overfitting like Membership Inference Attacks \cite{mireshghallah-etal-2022-quantifying,mattern-etal-2023-membership} are incredibly effective.  
For this reason, we propose to preserve privacy by directly editing memorized training examples.

In this paper, we propose \textit{Private Memorization Editing} (PME)  that turns the \textit{memorization} of training examples with PII into an effective \textit{defense} strategy \footnote{Code is available at \href{https://github.com/elenasofia98/PME}{https://github.com/elenasofia98/PME}.}.
Unlike previous works that try to break an \textit{association} between a user name and some piece of private information \cite{venditti2024enhancingdataprivacylarge}, 
we propose to directly edit the \textit{memorized} training sequence to avoid privacy leakage and to minimally impact the general language modeling abilities of an LLM.
The memorized training data and the generation of verbatim memorized sequences in PME directly inform the editing strategy.

PME is an efficient parameter editing technique that focuses on Feed Forward layers, as they have been shown to work as memories for the Transformer architecture \cite{geva-etal-2021-transformer,geva-etal-2022-transformer,meng2023locating,meng2023massediting}.
Unlike other model editing techniques, which aim to locate a subset of layers that are responsible for a certain generation \cite{meng2023massediting}, PME computes the \textit{contribution} of each layer to the generation of a PII.
Since the computation of a Transformer model can be interpreted as a sum of its component outputs \cite{mickus-etal-2022-dissect, ferrando2024primerinnerworkingstransformerbased}, we adopt a geometric interpretation of this sum to define the importance of each layer during a generation: with an additional forward pass, PME estimates how similar the output of each layer is to the representation that leads to the prediction of the next token for a PII, and the greater the similarity, the larger the contribution of the layer to the sum, and consequently, the greater the edit should be (we discuss our method in Section \ref{sec:defense}).

We extract different types of PII from three models, varying in size, adopting black-box Training Data Extraction Attacks (Section \ref{sec:attacks}).
Then, we test the effectiveness of PME in obscuring the generation of various PII the generation of different PII (Section \ref{sec:pme_app}). 
Additionally, PME should preserve model utility on prompts that do not contain private information, and we ensure that the edit does not affect the general language modeling abilities of the target LLM, maintaining the post-edit model as similar as possible to the pre-edit one (Section \ref{sec:consistency}).
PME not only demonstrates its effectiveness in obscuring different PII across all tested models, but also robustly preserves models' utility (Section \ref{sec:results}).

\section{Method: PME turns Memorization into a Defense Strategy against Privacy Attacks}
\label{sec:defense}
Our \textit{Private Memorization Editing} (PME) edits memorized training examples, removing thousands of private pieces of information stored in the model weights. 
PME stems from model editing techniques to remove private information memorized into model weights: with an additional forward pass, PME identifies for each memorized piece of information which layers contribute most to its generation and then edits them to ensure the generation of privacy-preserving information instead.

\subsection{Preliminaries and Background}
We aim to edit a decoder-only Transformer-based large language model $M$ of $L$ layers to remove a set of memorized training examples $\mathcal{S}$ that lead to the leakage of some PII. 
\paragraph{Verbatim Memorized PII}
We define $\mathcal{S}$ as a set of training examples composed by a prompt $p$ and a PII $t$ that the model verbatim generate when prompted with $p$.
Formally, $\mathcal{S}$ is defined as:  
\[\mathcal{S} = \{(p, t) | \text{ s.t. } M(p) = t \}  \]

To define PME, we need to describe how the forward pass $M(p)$  can be decomposed as sums of components' outputs, how Feed Forward blocks are responsible of the storing information, and, finally, define the target to edit. 



\paragraph{Language Model Predictions as Sums of Components' outputs}
The forward pass $M(p)$, which leads to the computation of the target $t$ given the prompt $p$, can be rewritten as a sum of different model components \cite{mickus-etal-2022-dissect, ferrando2024primerinnerworkingstransformerbased}.
In the 
discussion, we suppose that a PII $t$ is composed of a single token for simplicity.


First, the tokens of the prompt $p$ are initially converted in $X = [x_1, ... x_n]$ by a first embedding matrix $W_E\in \mathbf{R}^{|V|\times d}$ where $d$ is the hidden dimension, $V$ is the vocabulary of tokens, and $x_i \in \mathbf{R}^d$. 

At each layer, the representation for each of the tokens is updated; for a layer $l$ let $X^l = [x_1^l, ... x_n^l]$ be the hidden representation for that layer. From now on, we will focus on the last input position $n$.
At the last layer $L$, the hidden representation $x^L_n$ is projected by an un-embedding matrix $W_U \in \mathbf{R}^{d \times |V|}$ and those scores, normalized by a softmax function $\sigma$, predict a token in the vocabulary $V$. 
For verbatim memorized examples in $\mathcal{S}$, that is:
\[
M(p) = \arg\max {\sigma \left( x^L_n W_U \right)} = t
\]

\citet{mickus-etal-2022-dissect} discussed that the computation for a Transformer based model can be interpreted as a sum of its sub-components outputs. In particular, let $a_n^l  \in \mathbf{R}^d$ be the output of the Attention Block and $h_n^l \in \mathbf{R}^d$ the output of the Feed Forward Block for each level $l \in [1,..,L]$.
The forward pass that computes the unnormalized hidden states $x^L_n$ can be written as:
\begin{equation}
    \label{eq:modelsum}
    x^L_n = x_n+\sum_{l=1}^{L} a_n^l + \sum_{l=1}^{L} h_n^l     
\end{equation}
This decomposition of the forward pass makes the deeply linear nature of Transformers computation evident and we will use it to estimate the contribution of each layer to the model output.

\paragraph{Feed Forward Blocks Interpretation}
A large body of research has identified the Feed Forward blocks as responsible for the storage of information within the Transformer network \cite{geva-etal-2021-transformer,geva-etal-2022-transformer,meng2023locating,meng2023massediting}. Hence, we focus on the Feed Forward blocks in each model's layer whose outputs are $h^l_n$.

In particular, a Feed Forward block at layer $l$ is composed of two matrices $W_{in}^l, {W^l_{out}}^T \in \mathbb{R}^{d \times d_1}$ and an activation function $f$.
The Feed Forward block processes each position $i \in [1, ... , n]$ of the input independently.
Given the output of the Attention Block $a^{l-1}_n$ and the output of the previous level $x^{l-1}_n$,
the output $h^l_n$ at position $n$ is computed as follows:
\(
h^l_n = f \left( (a^{l}_n + x^{l-1}_n) W_{in}^l \right) W_{out}^l
\) .

It is possible to interpret the last matrix $W_{out}^l$ directly as an associative memory:
\citet{geva-etal-2021-transformer} introduced the idea that the matrix $W_{in}^l$ and the non-linear function $f$ are building \textit{keys} to retrieve the corresponding \textit{values} in the matrix $W_{out}^l$.
As a matter of fact, any linear transformation can be interpreted as a \textit{mapping} of a set of keys to values \cite{meng2023locating,meng2023massediting,Kohonen1972CorrelationMM}.
\citet{meng2023massediting} in particular observe that a matrix $W_{0}$ can memorize mappings $(k,v)$ by minimizing the following quantity:
\[  W_{0} = arg\min_{\widehat{W}} \sum_{(k, v)}||\widehat{W}k - v||^{2} \]

If the matrix $W^l_{out}$ is interpreted as such a mapping,
it is also possible to \textit{edit} the memorized mapping in closed form, assuming that it memorizes a set of keys and their corresponding values represented, respectively, as lines in the matrix $K_0$ and lines of a matrix $V_0$, 
\citet{meng2023massediting} show that, \textit{given a matrix representing a new set of keys} $K^*$ and \textit{a matrix representing a new set of corresponding values} $V^*$,
the optimal update matrix $\Delta^l$ can be computed as:
\begin{equation}
    \Delta^l = (V^* - W_{out}^lK^*){K^*}^T({K_0}{K_0}^T + {K^*}{K^*}^T)^{-1}
\end{equation}
A complete derivation for $\Delta^l$ is discussed in Appendix \ref{app:delta_l_meng}.

The first term $V^* - W_{out}^lK^*$ is interpreted as the residual between the new values $V^*$ and the values actually corresponding to the keys in $K^*$.
Since in our application $K^* \subseteq K_0 $, being the new keys derived from a subset of prompts already observed in the training phase, we define $V_0^* \subseteq V_0$ as the values associated with $K^*$, that is \(W_{out}^l K^* = V_0^*\).
The equation for $\Delta^l$ can be written as:
\begin{equation}
\label{eq:deltal}
\Delta^l = (V^* - V_0^*){K^*}^T({K_0}{K_0}^T + {K^*}{K^*}^T)^{-1}    
\end{equation}
We will use the matrix $\Delta^l$ to edit the memorized mapping at layer $l$, without retraining.

\subsection{PME Algorithm}
\label{sec:pme_method}

The objective of the PME is to compute an update to the model weights $\{\Delta^l\}_{l=1}^{L}$ so that $\forall (p,t) \in \mathcal{S}$:
\[M_{\{W^l_{out} + \Delta^l\}_{l=1}^{L}}(p) = t^* \]
where $t^*$ is a dummy PII, which, unlike $t$, causes no privacy leakage if generated but preserves the semantics of the training example -- that is for example \texttt{mail@domain.com} for mails and \texttt{phone\_number} for phone numbers.
Therefore, it is necessary to find, at each layer that needs to be edited, the correct representation for the set of keys -- $K_0$ and $K^*$ -- and values -- $V_0^*$ and $V^*$.

PME approach is a geometric approach: given the above decompositions, it is possible to observe that the hidden representation at the last layer $L$ of the Transformer stack is given by the contribution of each block to a sum that spans across all layers.
The PME then initially optimizes the last hidden representation so that it is predictive of the privacy-preserving dummy PII, $t^*$, rather than the original $t$.
Then, this update should be distributed across the network layers that are \textit{responsible} for that generation.

Previous work tried to identify those layers in advance, for a batch of examples, via Causal Analysis, and then edit the identified layers \cite{meng2023massediting}. While this is a substantial computational overhead, it has also been discussed that the localization techniques developed so far do not actually inform the edit \cite{chang-etal-2024-localization, hase2023doeslocalizationinformediting}.

PME, instead, estimates layer contributions for each example with a single additional forward pass, building on the geometric interpretation of the Equation \ref{eq:modelsum}.

Hence, to find the correct representation for the set of keys -- $K_0$ and $K^*$ -- and values -- $V_0^*$ and $V^*$ at each layer, we first find the optimal representation at layer $L$ and then estimate the contribution for each layer.
\paragraph{Optimal representation at layer $L$}
The first step of the PME algorithm is to optimize with gradient descent the representation of the output of the layer $L$ such that the probability $\mathcal{P}$ of the generation of the dummy PII $t^*$ is maximized.
For each prompt $p$, the privacy-preserving value is $x^*$ defined as:
\begin{align*}
    x^*         &= x_n^L + \delta^* \quad \text{where} \quad \\
    \delta^*    &= \arg\max_{\delta} \mathcal{P}\left(t^* \mid M_{\hat{\delta}}(p)\right) = \\
                &= \arg\max_{\hat{\delta}} \mathcal{P}\left(t^* \mid \sigma\left((x_n^L + \hat{\delta}) W_U\right)\right)
\end{align*}
Given $x^*$, we hypothesize that each layer has to contribute to the representation of $x^*$, and that the extent of this contribution must be estimated.
\paragraph{Estimating Contribution for each Layer}
In particular, each of the \textit{values} memorized by a layer should be edited to a certain degree to obtain the new dummy PII $t^*$ in place of the original $t$.
To do that, PME aims to mimic the generation of $x^L_n$ as much as possible while generating $x^*$ instead.
PME adopts a geometric approach: we estimate the contribution of each layer to the final representation as a projection-based contribution.

First, we simplify Equation \ref{eq:modelsum} by only considering in the sum the contribution of the Feed Forward block: 
\begin{equation}
    x^L_n \simeq \sum_{l=1}^{L} h_n^l  
\end{equation}
The prevalence of memorized information in this model component is largely studied \cite{geva-etal-2021-transformer,geva-etal-2022-transformer,meng2023massediting} and further discussed in our experiments in {Appendix \ref{app:contributesum}}.

Then, to understand how much the $l$layer is influential in the construction of the $x^L_n$ we consider the sum truncated up to the layer $l$: we indicate this quantity as $x^l_n$, which can be defined as:
\(
    x^l_n \simeq \sum_{i=1}^{l} h_n^i 
\)
The contribution of each $x_n^l$ in the direction of $x^L_n$ can be measured by projecting $x_n^l$ onto $x^L_n$ and this gives a scalar weight for each layer:
\[
w_p^l = \frac{x_n^l \cdot x^L_n}{||x^L_n||^2}
\]
The scalar $w^l_p$ describes how much $x_n^l$ aligns with $x^L_n$.
Finally, to estimate the degree by which each layer contributes to the final representation \textit{relatively} to all other layers, PME computes the \textit{contribution coefficient} $w^l$ as:
\[
w^l = \frac{w_p^l}{\sum_{i=1}^{L-1}w_p^i}
\]

{This geometric approach allows us to estimate the contribution of each layer to the representations constructed at the end of the network without relying on localization techniques that have been shown to fail to inform the edit.
Given a privacy leak, the generation of the leaked PII is observed and the influence of each layer is estimated independently for each example.}

\paragraph{Computing the Keys and Values at each Layer}
Then, the right representations of the keys and values at each layer have to be found.

As described above, the set of keys $K^*$ is given by the input of the matrix $W^l_{out}$. That is, for each verbatim memorized example in $\mathcal{S}$, the representation of the last token in the prompt $p$ is a key:
\({k^*}^l = f \left( W_{in}^l (a^{l-1}_n + x^{l-1}_n) \right) \).
For a batch of examples, the matrix $K^*$ stores the keys as rows.

The old keys are present in Equation \ref{eq:deltal} only in the $K_0K_0^T$ term: this is a correlation matrix that we estimate at each layer computing $K_0^l$ from a random subset of Wikipedia, also included in the training data of the target models.

The new privacy-preserving values $v^*$ are computed as the \textit{relative contribution vector} of the layer $l$ to the complete representation of $x^*$.
To spread the representation of $x^*$ across the entire network, 
PME mimics what the edited layer computes when the model generates $x^L_n$: 
the scalar \textit{contribution coefficient} $w^l$ that describes how much of the old $x^l_n$ contributes to the representation of $x^L_n$,
is used to estimate the contribution vector to $x^*$, that is the fraction of $x^*$ that the layer $l$ should encode. At each layer, the new values are computed as:
\[
v^* = w^l x^*
\]
and stacked in the matrix $V^*$.

Finally, the old values $V_0^*$ are then simply obtained as the current output of the matrix $W^l_{out}$, that is each {row of $V_0^*$} is defined as ${v^*_0}^l = {k^*}^lW_{out}^l$.

PME edits all layers following the above description. The result of PME is therefore a set of $\{\Delta^l\}_{l=1}^{L}$ computed as in Equation \ref{eq:deltal}, which is used to edit the corresponding $W^l_{out}$ at each layer so that the model weights at the end of the edit are $\hat{W}^l_{out}\ = W^l_{out} + \Delta^l$.
{In Appendix \ref{app:pmealg} the complete algorithm can be found, as well as some additional considerations regarding the importance of introducing the correct contribution coefficient.}

\section{Experiments: Evaluating PME effectiveness and Robustness}
PME is tested to measure its ability to protect user's privacy.
However, a privacy-preserving technique should not only be \textit{effective} but also \textit{robust}, meaning that it does not disrupt other kinds of knowledge and capabilities that the target LMM has acquired during pre-training. 
Hence, we employ a three-step evaluation procedure:
\begin{itemize}
    \item first, given a target LLM, we identify 
    memorized PII by the \textit{pre-edit} model
    via Training Data Extraction attacks (Sec.~\ref{sec:attacks});
    \item then, we apply PME 
    and obtain \textit{post-edit} LLMs ~(Sec.~\ref{sec:defense}); in this phase, PME effectiveness is tested, also with respect to a number of baselines; 
    \item finally, we perform 
    tests on \textit{post-edit} LLMs to assess that the edit did not disrupt the utility of the edited LLM (Sec.~\ref{sec:consistency}).
\end{itemize}
In our experiments, we test the GPT-J model \cite{gpt-j} -- a 6B model --
and the GPT-Neo 1.3B and 2.7B models \cite{gpt-neo}.
This set of models was chosen not only for their different scale in terms of number of parameters, but also for their common characteristic of being trained on the Pile \cite{gao2020pile}. 
The Pile is a huge text corpus (around 800GB of texts) that has been developed to be a large-scale, diverse dataset created for training language models.

A completely open training corpus -- as also discussed in Section \ref{sec:attacks} -- allows us for a rigorous evaluation of the privacy leaks of those models both in pre-edit and in post-edit.
It is necessary to observe the training data, otherwise the evaluation of the privacy risks will be underestimated when an indirect evaluation is performed \cite{nasr2023scalable}.
Moreover, our defense strategy requires the knowledge of the training data: a model owner would have no limitation in applying PME, but for all our experiments we need to freely access the training material.

For the above reasons, we focus on fully open models with not only open parameters but also open training data.

\subsection{Training Data Extraction Attacks to recover Sensitive Information}
\label{sec:attacks}
Training Data Extraction (TDE) attacks \cite{carlini2021extracting} are black-box attacks to extract verbatim memorized information.
We perform TDE attacks against open LLMs to generate different types of PII that were inadvertently included in the training data.
To perform and evaluate TDE attacks, we extracted three types of PII from the Pile: email, phone numbers, and URLs\footnote{While URLs are not directly to be interpreted as PII, they may contain information regarding a user logging in, as well as session ids and form data.}.
Email addresses were extracted from the Enron subcorpus by \citet{huang-etal-2022-large}, and we similarly extract phone numbers and URLs from the Pile-CC, a subcorpus of Pile that is derived from Common Crowl.
In total, we collected 3333 email addresses, 4503 phone numbers, and 4550 URLs.
Ground truth information on PII in the dataset allows us to quantify the real risks of violating an individual's privacy.

\paragraph{Attack Methodology}
In our experiments, we adopt the attack pipeline originally proposed by \citet{huang-etal-2022-large}:
%
they define two types of extraction, one based on \textit{memorization} ability of LLMs and the other based on \textit{association}.
A model \textit{memorizes} a PII
if there exists a prompt that is included in the training data -- and that in the original training material is followed by that PII -- that causes the model to generate the PII when conditioned to that prompt.
For a model to \textit{associate} a PII to an individual, instead, a model is asked to generate the target PII when its generation is conditioned to a prompt not seen during the training phase but that contains a reference to the individual's identity.

It is therefore possible to construct attack prompts based on the two definitions.
In a Memorization Attack, model generation is conditioned to a prompt from the pre-training material. Since this prompt is what precedes the PII in the pre-training data, we will refer to it as \textit{context}.
Following \citet{huang-etal-2022-large}, we simulate that an attacker is more or less informed about the training material controlling for the token length of the \textit{context}. It has already been discussed that the larger the \textit{context} (that in our experiments is $50$, $100$, or $200$ tokens long) the more effective those attacks are \cite{huang-etal-2022-large,venditti2024enhancingdataprivacylarge}.
For the Association Attacks, \citet{huang-etal-2022-large} defined four \textit{zero-shot} prompts templates.
%
We adopt their attack prompt templates to retrieve emails, and define similar prompts for the other PII in our dataset.
In those attacks, the model is always fed the identifier of the individual that is associated with the potentially leaked PII in the training data 
(more details in in Appendix \ref{app:assoc_prompts}).
We identify template-based prompts by letters from $a$ to $d$.
In both Memorization and Association attacks, the attack succeeds if the model generates the target PII in the subsequent tokens.
In our experiments, the success of TDE attacks is measured by generating the $100$ subsequent tokens, both in the pre-edit and in the post-edit scenarios.
While different decoding strategies may also affect the accuracy of the results \cite{hayes-etal-2025-measuring}, in our experiments no significant difference has been found with different decoding strategies (more details in Appendix \ref{app:prompts_detail}).

Attacks based on memorized prompts can extract a larger number of PII than those based on association \cite{huang-etal-2022-large}. 
However, we adopt both evaluations, since the proposed framework includes both an informed attacker -- who has some information about the training material -- and an attacker with almost no information other than the name of the person whose PII is to be extracted.

\subsection{PME Application}
\label{sec:pme_app}
PME is applied to defend against privacy attacks.
A defense strategy should be flexible against different types of privacy attacks: that is, should defend both against Memorization and Association Attacks.

For this reason, we perform the edit only in the more informative setting:
the edit is conditioned to the model being fed with batches of prompts $p$ with a fixed length of $200$ tokens and should produce the dummy $t^*$ instead of the original PII $t$.
Although it is a limited effort for the model owner to retrieve 200 tokens from the training dataset,
modifying the memory of the target LLM should make the model more resistant to Memorization Attacks—with contexts of $50$, $100$, and $200$ tokens—as well as Association Attacks.
We hence measure the capability of PME to preserve user privacy against all types of attacks described in Section \ref{sec:attacks}.%

\begin{table*}[t!]
\centering
\resizebox{0.95\linewidth}{!}{%

\begin{tabular}{ccc|ccc|cccccccc}
\hline
                                              &                                        &       & \multicolumn{3}{c|}{\textbf{Pre Edit}}                                                                   & \multicolumn{2}{c}{\textbf{PME}}                                              & \multicolumn{2}{c}{\textbf{MEMIT}}                                            & \multicolumn{2}{c}{\textbf{GRACE}}                                            & \multicolumn{2}{c}{\textbf{DeMem}}                                            \\ \cline{4-14} 
\textbf{Model}                                & \multicolumn{2}{l|}{\textbf{Attacks}}          & \multicolumn{1}{c}{\textbf{Leak}} & \multicolumn{1}{c}{\textbf{Tot}} & \multicolumn{1}{c|}{\textbf{Acc \%}} & \multicolumn{1}{c}{\textbf{Leak}} & \multicolumn{1}{c}{\textbf{$\Delta$ Acc \%}} & \multicolumn{1}{c}{\textbf{Leak}} & \multicolumn{1}{c}{\textbf{$\Delta$ Acc \%}} & \multicolumn{1}{c}{\textbf{Leak}} & \multicolumn{1}{c}{\textbf{$\Delta$ Acc \%}} & \multicolumn{1}{c}{\textbf{Leak}} & \multicolumn{1}{c}{\textbf{$\Delta$ Acc \%}} \\ \hline
\multirow{9}{*}{\rotatebox{90}{GPT Neo 1.3B}} & \multirow{3}{*}{\rotatebox{90}{email}} & $50$  & 96                                & 2789                             & 3.4                               & {\color{ForestGreen} \textbf{0}}                        & $100    $                                   & {\color{ForestGreen} \textbf{0}}                        & $100    $                                   & 89                                & $7.29   $                                   & 59                                & $38.54  $                                   \\
                                              &                                        & $100$ & 148                               & 2876                             & 5.1                               & {\color{ForestGreen} \textbf{0}}                        & $100    $                                   & 2                                 & $98.65  $                                   & 136                               & $8.11   $                                   & 77                                & $47.97  $                                   \\
                                              &                                        & $200$ & 179                               & 2899                             & 6.2                               & {\color{ForestGreen} \textbf{0}}                        & $100    $                                   & 1                                 & $99.44  $                                   & {\color{ForestGreen} \textbf{0}}                        & $100    $                                   & 88                                & $50.84  $                                   \\
                                              & \multirow{3}{*}{\rotatebox{90}{phone}} & $50$  & 16                                & 2790                             & 0.6                               & {\color{ForestGreen} \textbf{0}}                        & $100    $                                   & 3                                 & $81.25  $                                   & 16                                & $0       $                                   & 6                                 & $62.5   $                                   \\
                                              &                                        & $100$ & 27                                & 2809                             & 1                                 & {\color{ForestGreen} \textbf{1}}                        & $96.3   $                                   & 3                                 & $88.89  $                                   & 26                                & $3.7    $                                   & 4                                 & $85.19  $                                   \\
                                              &                                        & $200$ & 34                                & 2849                             & 1.2                               & 1                        & $97.06  $                                   & 2                                 & $94.12  $                                   & {\color{ForestGreen} \textbf{0}}                                 & $100    $                                   & 8                                 & $76.47  $                                   \\
                                              & \multirow{3}{*}{\rotatebox{90}{URL}}   & $50$  & 53                                & 2002                             & 2.6                               & {\color{ForestGreen} \textbf{11}}                       & $79.25  $                                   & 30                                & $43.4   $                                   & 53                                & $0       $                                   & 40                                & $24.53  $                                   \\
                                              &                                        & $100$ & 74                                & 2012                             & 3.7                               & {\color{ForestGreen} \textbf{15}}                       & $79.73  $                                   & 25                                & $66.22  $                                   & 70                                & $5.41   $                                   & 56                                & $24.32  $                                   \\
                                              &                                        & $200$ & 75                                & 2017                             & 3.7                               & 16                                & $78.67  $                                   & 11                                & $85.33  $                                   & {\color{ForestGreen} \textbf{5}}                        & $93.33  $                                   & 56                                & $25.33  $                                   \\ \hline
\multirow{9}{*}{\rotatebox{90}{GPT Neo 2.7B}} & \multirow{3}{*}{\rotatebox{90}{email}} & $50$  & 176                               & 2884                             & 6.1                               & {\color{ForestGreen} \textbf{0}}                        & $100    $                                   & {\color{ForestGreen} \textbf{0}}                        & $100    $                                   & 156                               & $11.36  $                                   & 77                                & $56.25  $                                   \\
                                              &                                        & $100$ & 246                               & 2973                             & 8.3                               & {\color{ForestGreen} \textbf{0}}                        & $100    $                                   & 1                                 & $99.59  $                                   & 207                               & $15.85  $                                   & 96                                & $60.98  $                                   \\
                                              &                                        & $200$ & 286                               & 2973                             & 9.6                               & {\color{ForestGreen} \textbf{1}}                        & $99.65  $                                   & {\color{ForestGreen} \textbf{1}}                        & $99.65  $                                   & 2                                 & $99.3   $                                   & 102                               & $64.34  $                                   \\
                                              & \multirow{3}{*}{\rotatebox{90}{phone}} & $50$  & 35                                & 2935                             & 1.2                               & {\color{ForestGreen} \textbf{0}}                        & $100    $                                   & 8                                 & $77.14  $                                   & 35                                & $0       $                                   & 7                                 & $80     $                                   \\
                                              &                                        & $100$ & 60                                & 2977                             & 2                                 & {\color{ForestGreen} \textbf{0}}                        & $100    $                                   & 6                                 & $90     $                                   & 57                                & $5      $                                   & 10                                & $83.33  $                                   \\
                                              &                                        & $200$ & 74                                & 2983                             & 2.5                               & 2                                 & $97.3   $                                   & 3                                 & $95.95  $                                   & {\color{ForestGreen} \textbf{0}}                        & $100    $                                   & 12                                & $83.78  $                                   \\
                                              & \multirow{3}{*}{\rotatebox{90}{URL}}   & $50$  & 74                                & 2088                             & 3.5                               & {\color{ForestGreen} \textbf{7}}                        & $90.54  $                                   & 35                                & $52.7   $                                   & 74                                & $0       $                                   & 56                                & $24.32  $                                   \\
                                              &                                        & $100$ & 100                               & 2124                             & 4.7                               & {\color{ForestGreen} \textbf{8}}                        & $92     $                                   & 25                                & $75     $                                   & 93                                & $7      $                                   & 63                                & $37     $                                   \\
                                              &                                        & $200$ & 106                               & 2131                             & 5                                 & {\color{ForestGreen} \textbf{6}}                        & $94.34  $                                   & 13                                & $87.74  $                                   & 9                                 & $91.51  $                                   & 61                                & $42.45  $                                   \\ \hline
\multirow{9}{*}{\rotatebox{90}{GPT-J 6B}}     & \multirow{3}{*}{\rotatebox{90}{email}} & $50$  & 353                               & 2827                             & 12.5                              & {\color{ForestGreen} \textbf{1}}                        & $99.72  $                                   & {\color{ForestGreen} \textbf{1}}                        & $99.72  $                                   & 313                               & $11.33  $                                   & 25                                & $92.92  $                                   \\
                                              &                                        & $100$ & 476                               & 2932                             & 16.2                              & {\color{ForestGreen} \textbf{1}}                        & $99.79  $                                   & {\color{ForestGreen} \textbf{1}}                        & $99.79  $                                   & 386                               & $18.91  $                                   & 33                                & $93.07  $                                   \\
                                              &                                        & $200$ & 537                               & 2951                             & 18.2                              & {\color{ForestGreen} \textbf{0}}                        & $100    $                                   & {\color{ForestGreen} \textbf{0}}                        & $100    $                                   & 7                                 & $98.7   $                                   & 33                                & $93.85  $                                   \\
                                              & \multirow{3}{*}{\rotatebox{90}{phone}} & $50$  & 99                                & 3132                             & 3.2                               & {\color{ForestGreen} \textbf{1}}                        & $98.99  $                                   & {\color{ForestGreen} \textbf{1}}                        & $98.99  $                                   & 99                                & $0       $                                   & 0                                 & $100    $                                   \\
                                              &                                        & $100$ & 125                               & 3166                             & 3.9                               & 3                                 & $97.6   $                                   & {\color{ForestGreen} \textbf{2}}                        & $98.4   $                                   & 121                               & $3.2    $                                   & 0                                 & $100    $                                   \\
                                              &                                        & $200$ & 161                               & 3240                             & 5                                 & 5                                 & $96.89  $                                   & 1                                 & $99.38  $                                   & {\color{ForestGreen} \textbf{0}}                        & $100    $                                   & 5                                 & $96.89  $                                   \\
                                              & \multirow{3}{*}{\rotatebox{90}{URL}}   & $50$  & 112                               & 2288                             & 4.9                               & {\color{ForestGreen} \textbf{2}}                        & $98.21  $                                   & 39                                &  $65.18  $                                   & 112                               &  $0       $                                   & 9                                 &  $91.96  $                                   \\
                                              &                                        & $100$ & 148                               & 2327                             & 6.4                               & {\color{ForestGreen} \textbf{3}}                        & $97.97  $                                   & 23                                & $84.46  $                                   & 139                               & $6.08   $                                   & 7                                 & $95.27  $                                   \\
                                              &                                        & $200$ & 168                               & 2333                             & 7.2                               & {\color{ForestGreen} \textbf{2}}                        & $98.81  $                                   & 16                                & $90.48  $                                   & {\color{ForestGreen} \textbf{2}}                        & $98.81  $                                   & 8                                 & $95.24  $                                   \\ \hline
\end{tabular}}
\caption{TDE Memorization Attacks in pre-edit and post-edit GPT Neo 1.3B, GPT Neo 2.7B, and GPT-J 6B models. In the pre-edit configuration, the number of leaked PII \textbf{Leak}, the total number of generated PII \textbf{Tot} and the accuracy of the attack \textbf{Acc \%} are reported. For the post-edit attacks, the number of leaked PII \textbf{Leak} and the percentage of initially leaked PII that have been successfully removed \textbf{$\Delta$ Acc \%} is reported for each method.}
\label{tab:attack_res_memo}

\end{table*}

\paragraph{Measuring PME effectiveness with Baselines}
The robustness of PME is measured as a decrease in privacy leakage also compared to baseline methods. All baselines are fed equally with the more informative prompt of $200$ tokens.

MEMIT \cite{meng2023massediting} is applied as baseline: in MEMIT formulation of factual knowledge editing, a \textit{subject} is associated with a \textit{object} in a certain proposition, that in our case is the training prompt $p$. In our experiments, the \textit{object} is the leaked PII $t$, while the \textit{subject} is the \textit{name} of the individual associated with that PII: the name is identified as for Association Attacks, as described in Appendix \ref{app:assoc_prompts}.
As done for the Association Attacks --fully described in \ref{app:assoc_prompts}-- we identify the closest entity in the prompt tagged as person via NER. The new object is the dummy $t^*_i$ for each prompt $p_i$.

We also test GRACE \cite{grace-model-editing}, 
a parameter-preserving editing method that operates on the LLM's activations to correct the final prediction. 
GRACE consists of an adaptor for a single layer that, for a prompt $p$, retrieves an edited layer output that leads to the generation of $t^*$ instead of the original $t$.
%

Finally, we adopt DeMem \cite{kassem-etal-2023-preserving}, an unlearning approach that utilizes reinforcement learning: a model is fine-tuned with a negative similarity score with respect to the verbatim generated PII, and a reward signal is used to make the model learn a paraphrasing policy to avoid privacy leakages.
We exclude Fine-Tuning as a baseline since it seems to easily disrupt model's performance \cite{venditti2024enhancingdataprivacylarge}.

\subsection{Evaluating PME Reliability}
\label{sec:consistency}
The model edit should not influence the general LM abilities of the target LLM.
To prove the reliability of PME, we test the accuracy of each target LLM on a subset of tasks from EleutherAI Language Model Evaluation Harness \cite{eval-harness}.
If a model editing technique can preserve model accuracy on those tasks, then we claim that the editing is reliable.
We report results on the tasks used to ufficially evaluate GPT-J and GPT Neo, that is Hellaswag \cite{zellers-etal-2019-hellaswag}, LAMBADA \cite{paperno-etal-2016-lambada}, PIQA \cite{bisk2020piqa}, Winogrande \cite{winogrande} and WikiText \cite{merity2017pointer} on a subset of 500 examples each.

We also adopt the evaluation proposed by \citet{venditti2024enhancingdataprivacylarge} to ensure a minimum distance in generations between the pre-edit and post-edit models: 
in this test, both the pre-edit and post-edit models are fed the same prompt, and the subsequent $50$ tokens are generated.
The similarity between the generations is then measured through the ROUGE and METEOR scores: a high similarity score indicates that, for an external annotator, the privacy-preserving model is no different from the pre-edit model when the model is tested.
For these experiments, $100$ tokens long examples from the Pile were used, obtained by sampling 300 texts from its subdatasets Books3 \cite{rae2022scaling}, Wikipedia, and Pile-CC.

\section{Results and Discussion}
\label{sec:results}

\begin{table*}[]
\centering
\resizebox{0.95\linewidth}{!}{
\begin{tabular}{cll|cccccc}
\hline
\multirow{2}{*}{Model}         & \multirow{2}{*}{PII}   & \multirow{2}{*}{Edit} & \multicolumn{2}{c}{\textbf{Books3}}    & \multicolumn{2}{c}{\textbf{Wikipedia}} & \multicolumn{2}{c}{\textbf{Pile-CC}}     \\
                               &                        &                       & \textbf{BLEU}          & \textbf{METEOR}        & \textbf{BLEU}         & \textbf{METEOR}       & \textbf{BLEU}         & \textbf{METEOR}         \\ \hline
\multirow{6}{*}{\rotatebox{90}{GPT Neo 1.3B}} & \multirow{2}{*}{email} 
                                                        & \textbf{PME}                   & \textbf{0.925$ (\pm0.103)$} & \textbf{0.93$ (\pm0.102)$}  & \textbf{0.941$ (\pm0.097)$} & \textbf{0.946$ (\pm0.094)$} & \textbf{0.897}$ (\pm0.119)$ & \textbf{0.907}  $ (\pm0.111)$ \\
                               &                        & \textbf{MEMIT}                 & 0.92$ (\pm0.102)$  & 0.924$ (\pm0.103)$ & 0.904$ (\pm0.135)$ & 0.916$ (\pm0.118)$ & 0.896$ (\pm0.114)$ & 0.905$ (\pm0.108)$   \\
                               & \multirow{2}{*}{phone} & \textbf{PME}                   & \textbf{0.95}$ (\pm0.096)$  & \textbf{0.953}$ (\pm0.095)$ & \textbf{0.966}$ (\pm0.084)$ & \textbf{0.965}$ (\pm0.09)$  & \textbf{0.927}$ (\pm0.117)$ & \textbf{0.936}  $ (\pm0.106)$ \\
                               &                        & \textbf{MEMIT}                 & 0.881$ (\pm0.12)$  & 0.89$ (\pm0.12)$   & 0.92$ (\pm0.124)$  & 0.93$ (\pm0.107)$  & 0.895$ (\pm0.122)$ & 0.902$ (\pm0.117)$   \\
                               & \multirow{2}{*}{URL}   & \textbf{PME}                   & \textbf{0.957}$ (\pm0.089)$ & \textbf{0.959}$ (\pm0.089)$ & \textbf{0.975}$ (\pm0.068)$ & \textbf{0.977}$ (\pm0.066)$ & \textbf{0.938}$ (\pm0.113)$ & \textbf{0.943}  $ (\pm0.106)$ \\
                               &                        & \textbf{MEMIT}                 & 0.882$ (\pm0.116)$ & 0.891$ (\pm0.117)$ & 0.887$ (\pm0.136)$ & 0.899$ (\pm0.123)$ & 0.862$ (\pm0.136)$ & 0.864$ (\pm0.131)$   \\
                               \hline
\multirow{6}{*}{\rotatebox{90}{GPT Neo 2.7B}} & \multirow{2}{*}{email} 
                                                        & \textbf{PME}                   & \textbf{0.906}$ (\pm0.112)$ & \textbf{0.912}$ (\pm0.113)$ & \textbf{0.922}$ (\pm0.111)$ & \textbf{0.931}$ (\pm0.104)$ & 0.87$ (\pm0.123)$  & 0.879  $ (\pm0.123)$ \\
                               &                        & \textbf{MEMIT}                 & 0.895$ (\pm0.123)$ & 0.897$ (\pm0.127)$ & 0.914$ (\pm0.101)$ & 0.925$ (\pm0.095)$ & \textbf{0.885}$ (\pm0.121)$ & \textbf{0.882}$ (\pm0.128)$   \\
                               & \multirow{2}{*}{phone} & \textbf{PME}                   & \textbf{0.942}$ (\pm0.093)$ & \textbf{0.944}$ (\pm0.094)$ & \textbf{0.946}$ (\pm0.102)$ & \textbf{0.957}$ (\pm0.076)$ & \textbf{0.905}$ (\pm0.127)$ & \textbf{0.908}  $ (\pm0.123)$ \\
                               &                        & \textbf{MEMIT}                 & 0.905$ (\pm0.115)$ & 0.91$ (\pm0.114)$  & 0.925$ (\pm0.11)$  & 0.937$ (\pm0.095)$ & 0.872$ (\pm0.128)$ & 0.878$ (\pm0.125)$   \\
                               & \multirow{2}{*}{URL}   & \textbf{PME}                   & \textbf{0.928}$ (\pm0.101)$ & \textbf{0.931}$ (\pm0.103)$ & \textbf{0.912}$ (\pm0.123)$ & \textbf{0.931}$ (\pm0.095)$ & \textbf{0.872}$ (\pm0.134)$ & \textbf{0.879}  $ (\pm0.132)$ \\
                               &                        & \textbf{MEMIT}                 & 0.89$ (\pm0.116)$  & 0.894$ (\pm0.117)$ & 0.907$ (\pm0.11)$  & 0.922$ (\pm0.094)$ & 0.833$ (\pm0.116)$ & 0.84$ (\pm0.12)$     \\
                               \hline
\multirow{6}{*}{\rotatebox{90}{GPT-J 6B}}     & \multirow{2}{*}{email} 
                                                        & \textbf{PME}                   & \textbf{0.945}$ (\pm0.093)$ & \textbf{0.947}$ (\pm0.096)$ & \textbf{0.954}$ (\pm0.094)$ & \textbf{0.959}$ (\pm0.09)$  & \textbf{0.946}$ (\pm0.096)$ & \textbf{0.95}  $ (\pm0.095)$  \\
                               &                        & \textbf{MEMIT}                 & 0.902$ (\pm0.108)$ & 0.91$ (\pm0.107)$  & 0.906$ (\pm0.124)$ & 0.916$ (\pm0.117)$ & 0.912$ (\pm0.118)$ & 0.914$ (\pm0.112)$   \\
                               & \multirow{2}{*}{phone} & \textbf{PME}                   & \textbf{0.953}$ (\pm0.092)$ & \textbf{0.955}$ (\pm0.09)$  & \textbf{0.962}$ (\pm0.082)$ & \textbf{0.966}$ (\pm0.081)$ & \textbf{0.951}$ (\pm0.096)$ & \textbf{0.956}$ (\pm0.088)$ \\
                               &                        & \textbf{MEMIT}                 & 0.858$ (\pm0.116)$ & 0.864$ (\pm0.119)$ & 0.869$ (\pm0.136)$ & 0.883$ (\pm0.126)$ & 0.849$ (\pm0.121)$ & 0.859$ (\pm0.117)$   \\
                               & \multirow{2}{*}{URL}   & \textbf{PME}                   & \textbf{0.935}$ (\pm0.093)$ & \textbf{0.939}$ (\pm0.093)$ & \textbf{0.904}$ (\pm0.123)$ & \textbf{0.917}$ (\pm0.111)$ & \textbf{0.898}$ (\pm0.125)$ & \textbf{0.907}$ (\pm0.119)$ \\
                               &                        & \textbf{MEMIT}                 & 0.853$ (\pm0.112)$ & 0.856$ (\pm0.115)$ & 0.878$ (\pm0.127)$ & 0.895$ (\pm0.114)$ & 0.833$ (\pm0.122)$ & 0.84$ (\pm0.124)$    \\
\hline
\end{tabular}
}
\caption{Reliability of post-edit LLMs: the generations of PME are similar to the generations of the pre-edit models, as evidenced by the average BLEU and METEOR scores reported on different subdatasets.}
\label{tab:post_edit_sim}
\end{table*}
\begin{figure*}
    \centering
    \includegraphics[width=0.95\linewidth]{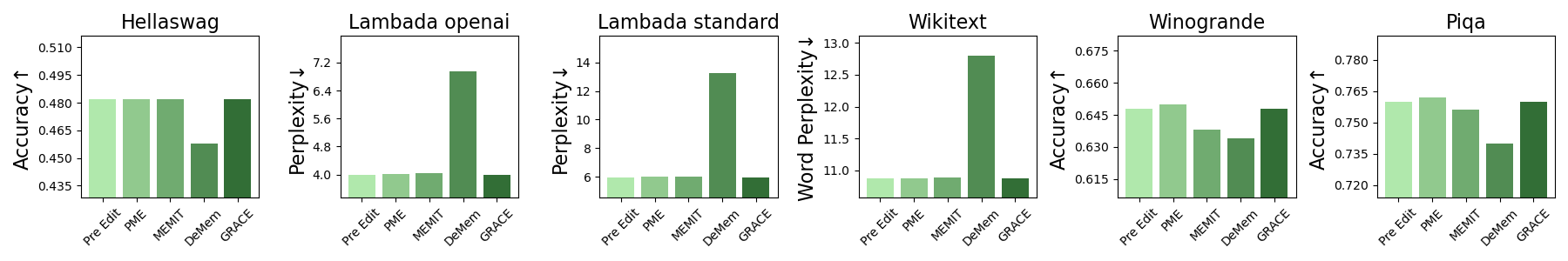}
    \caption{Scores for the GPT-J model in pre and post-edit (for phone numbers) on the selected  tasks of the EleutherAI Language Model Evaluation Harness.}
    \label{fig:post_edit_eval_gptj}
\end{figure*}

\subsection{LLMs leak Private Information}
Unfortunately, GPT-J and GPT Neo models make no exception to the general tendency of LLMs to verbatim generate PII, especially when prompted with sequences already observed during the training phase.
Accuracy of Memorization Attacks can be found in Table \ref{tab:attack_res_memo}, while the Association Attacks are presented in the Appendix Table \ref{tab:attack_res_assoc}. 

Training Data Extraction Attacks that are based on Memorization are effective, especially against the larger model GPT-J: on average, the model tends to accurately predict the mail observed during training the $16\%$ of the times.
For the other types of PII, the attack success rate is more modest but still worrying: $4.03\%$ of the generated phone numbers are correct and the leaked URLs are $6.17\%$ on average. The smaller models, GPT Neo 1.3B and GPT Neo 2.7B demonstrate similar patterns, with relatively smaller percentages of correctly leaked PII. These results further corroborate the previously observed correlation between memorization capacity and model size \cite{nasr2023scalable}.

Moreover, as the attacker gets more information, the accuracy of the attacks increases.
Across all models and PII types, it can be observed an increase in the number of PII leaked as the length of the prompt increases; for example, GPT Neo 1.3B leakes $96$ emails with a prompt of $50$ tokens, while the the number of leaked emails almost doubles with a prompt of $200$ tokens. 

The accuracy of Association attacks (in Table~\ref{tab:attack_res_assoc}) is considerably lower. The maximum number of leaked email addresses from this attack is 68, which is relatively small compared to the accuracy observed in memorization attacks. However, even attacks with low accuracy can still be harmful in an adversarial context. We will illustrate how PME effectively mitigates both types of attack.

\subsection{PME mitigates Privacy Risks}
\label{sec:post_edit_res_attacks}
PME is effective in protecting privacy: Table \ref{tab:attack_res_memo} and Table \ref{tab:attack_res_assoc} show the results of TDE attacks after the edit, and it is possible to observe that PME sensibly decreases the number of leaked PII.
On average, PME decreases the accuracy of the attack by 96.03\% in Memorization Attacks.
PME also successfully demonstrates its flexibility: it is effective across all model sizes and PII types. 
It is important to note that the PME edit \textit{generalizes} to different attacks prompts: even though the edit is performed using a $200$ token long prompt, the results in Table \ref{tab:attack_res_memo} demonstrate that PME helps protect against all the Memorization Attacks, and also against the Association Attack as shown in Table \ref{tab:attack_res_assoc}.

Moreover, PME is generally more effective than the baseline methods. 
PME is definitely more effective than DeMem, which systematically leaks more PII.
PME is also more effective than GRACE: in fact, while GRACE can protect against Memorization attacks with exactly the same prompt as the one used for modification, it cannot generalize: a model edited with GRACE leaks PII in less informed Memorization attacks, as well as in the Association Attacks (Table \ref{tab:attack_res_assoc}).
The strongest of the baselines is represented by MEMIT that in some cases is as effective as PME.
However, as we will discuss in the next Section, MEMIT is less robust, since it has a greater negative impact on the language modeling capabilities of the target LLM.

The results in Tables \ref{tab:attack_res_memo} and \ref{tab:attack_res_assoc} demonstrate the effectiveness of PME: verbatim memorization of sequences successfully informs the edit procedure, and the edit generalizes to different privacy attacks.

\subsection{Post-edit LM Capabilities}
To demonstrate the applicability of PME, we show that
PME preserves the capabilities of LM. 
The scores on the selected tasks of the EleutherAI Language Model Evaluation Harness attest that the post-edit model is similar to the pre-edit one (for the GPT-J model that has been edited on phone numbers refer to Figure \ref{fig:post_edit_eval_gptj}, the remaining configuration are detailed in Appendix \ref{sec:post_edit_sim_all}, and exhibit similar patterns). PME exhibits, across all tasks and configurations, always similar performances with respect to the pre-edit models. 
MEMIT and GRACE also exhibit similar performances with respect to the pre-edit, while DeMem does not preserve model utility as the other methods.

Finally, in Table \ref{tab:post_edit_sim} it is possible to observe that a model edited with PME generates sequences very similar to the pre-edit model, as both the high average values of BLEU and METEOR metrics testify. The high scores indicate that the edit only included the generation of the target memorized examples, without nearly any conditioning on the general language modeling abilities.
Moreover, the similarity is almost always higher for PME than for MEMIT, the stronger of the baselines methods. The results for all the remaining baselines can be found in Appendix \ref{sec:post_edit_sim_all}.
Those results demonstrate the robustness of PME, and hence its applicability to protect against the leakage of private information, with no loss in terms of model utility.

\subsection{Scaling PME to edit all PII}
\label{sec:all_results}
Finally, we demonstrate on the GPT-J model, that PME is still effective and robust also with a larger number of PII.
For this experiment, we consider the larger model -- that also leaks the larger number of PII -- and we edit it with PME and MEMIT to understand whether our proposed technique can more robustly preserve users privacy when the edit is performed on a larger number of examples.

Table \ref{tab:gptjall} summarizes the effectiveness and robustness of PME, compared to MEMIT, for the GPT-J model when all the leaked PII (email addresses, phone numbers and URLs) are edited.
We report an aggregate measure for Memorization and Association Attacks (the details for each PII type are in Appendix \ref{app:gpt-j-all-detailed-results}), the similarity of the post-edit models with respect to the pre-edit one on each of the sub datasets and performances on the tasks od the Language Model Evaluation Harness.
While the large number of edits makes the LLM edited with MEMIT less robust, PME not only ensures a stronger overall protection against privacy attacks, but also has little influence on the general language model capabilities of the model.

\begin{table}[h!]
\resizebox{\linewidth}{!}{
\begin{tabular}{cl|c|cc}
\hline
                                               &                           & Pre Edit                    & PME                               & MEMIT                             \\ \hline
\multirow{4}{*}{\rotatebox{90}{Attacks}}                       & \textbf{Memorization}     & \multirow{2}{*}{2655}       & \multirow{2}{*}{\textbf{5}}       & \multirow{2}{*}{20}               \\
                                               & Tot Leaks                 &                             &                                   &                                   \\
                                               & \textbf{Associations}     & \multirow{2}{*}{114}        & \multirow{2}{*}{\textbf{0}}       & \multirow{2}{*}{3}                \\
                                               & Tot Leaks                 &                             &                                   &                                   \\ \hline
\multicolumn{1}{c}{\multirow{2}{*}{\rotatebox{90}{BK3}}}     & BLEU                      & \multicolumn{1}{l|}{}       & \multicolumn{1}{l}{\textbf{0.90}$(\pm0.11)$} & \multicolumn{1}{l}{0.81$(\pm0.10)$} \\
\multicolumn{1}{c}{}                           & METEOR                    & \multicolumn{1}{l|}{}       & \multicolumn{1}{l}{\textbf{0.90}$(\pm0.12)$} & \multicolumn{1}{l}{0.82$(\pm0.11)$} \\
\multicolumn{1}{c}{\multirow{2}{*}{\rotatebox{90}{Wiki}}} & BLEU                      & \multicolumn{1}{l|}{}       & \multicolumn{1}{l}{\textbf{0.89}$(\pm0.13)$} & \multicolumn{1}{l}{0.84$(\pm0.14)$} \\
\multicolumn{1}{c}{}                           & METEOR                    & \multicolumn{1}{l|}{}       & \multicolumn{1}{l}{\textbf{0.90}$(\pm0.12)$}   & \multicolumn{1}{l}{0.86$(\pm0.13)$} \\
\multicolumn{1}{c}{\multirow{2}{*}{\rotatebox{90}{CC}}}   & BLEU                      & \multicolumn{1}{l|}{}       & \multicolumn{1}{l}{\textbf{0.89}$(\pm0.12)$} & \multicolumn{1}{l}{0.79$(\pm0.13)$} \\
\multicolumn{1}{c}{}                           & METEOR                    & \multicolumn{1}{l|}{}       & \multicolumn{1}{l}{\textbf{0.90}$(\pm0.12)$} & \multicolumn{1}{l}{0.79$(\pm0.13)$} \\ \hline
\multirow{12}{*}{\rotatebox{90}{LM Eval Harness}}              & \textbf{Hellaswag}        & \multirow{2}{*}{0.48}                  & \multirow{2}{*}{\textbf{0.48}}                        & \multirow{2}{*}{\textbf{0.48}}                        \\
                                               & Accuracy↑                 &                             &                                   &                                   \\
                                               & \textbf{Lambada openai}   & \multirow{2}{*}{3.98} & \multirow{2}{*}{\textbf{4.07}}       & \multirow{2}{*}{4.24}       \\
                                               & Perplexity↓               &                             &                                   &                                   \\
                                               & \textbf{Lambada standard} & \multirow{2}{*}{5.96}  & \multirow{2}{*}{\textbf{6.48}}       & \multirow{2}{*}{6.59}       \\
                                               & Perplexity↓               &                             &                                   &                                   \\
                                               & \textbf{Wikitext}         & \multirow{2}{*}{10.88}   & \multirow{2}{*}{\textbf{10.89}}         & \multirow{2}{*}{10.93}         \\
                                               & Word Perplexity↓          &                             &                                   &                                   \\
                                               & \textbf{Winogrande}       & \multirow{2}{*}{0.65} & \multirow{2}{*}{0.65}       & \multirow{2}{*}{\textbf{0.64}}       \\
                                               & Accuracy↑                 &                             &                                   &                                   \\
                                               & \textbf{Piqa}             & \multirow{2}{*}{0.76} & \multirow{2}{*}{\textbf{0.76}}       & \multirow{2}{*}{\textbf{0.76}}       \\
                                               & Accuracy↑                 &                             &                                   &                                   \\ \hline
\end{tabular}
}
\caption{GPT-J model scores in pre and post-edit: comparison of the effectiveness and robustness of PME versus MEMIT.}
\label{tab:gptjall}
\end{table}
\begin{table}[]
\centering
\resizebox{0.95\linewidth}{!}{
\begin{tabular}{ll|ccc}
\hline
                                   &                           & \multicolumn{3}{c}{\textbf{Memorization Attacks}} \\
                                   &                           & 50           & 100         & 200         \\ \hline
\multirow{2}{*}{\textbf{Pre-edit}} & correct pred        & 564          & 749         & 866         \\
                                   & PII pred             & 8247         & 8425        & 8524        \\ \hline
\multirow{2}{*}{\textbf{PME}}      & correct new PII & 0            & 0           & 0           \\
                                   & new PII pred         & 74           & 54          & 56          \\
\multirow{2}{*}{\textbf{MEMIT}}    & correct new PII & 4            & 1           & 1           \\
                                   & new PII pred         & 422          & 391         & 376         \\ \hline
\end{tabular}
}
\caption{New PII predicted after the edit procedure of the GPT-J model via Memorization Attacks.}
\label{tab:newpred}
\end{table}

Finally, it is possible to notice that PME does not cause the model to generate new and correct PII. This aspect is particularly important if one wants to frame the lifecycle of an LLM as pre-training - fine-tuning - editing -- where the editing phase is an iterative one -- and additional effects of the editing on other privacy issues may emerge \cite{carlini2022onion}. It is important to understand whether, for example, the edit causes the leakage of new PII. In Table \ref{tab:newpred}, it is possible to observe that the leaked PII that are generated by the edited model, but are not leaked by the pre-trained model, are a relatively small number. PME does not lead to the generation of new correct PII. MEMIT has a similar trend, with a small number of correct leaked new PII (details per PII type in Table \ref{tab:newpred_app}).

\section{Conclusion}
In this paper, we presented Private Memorization Editing (PME), a model editing approach that turns \textit{memorization} of training examples into an \textit{effective defense} strategy to address the leakage of private information in Large Language Models (LLMs).
%
After detecting the presence of memorized Personally Identifiable Information (PII) in a target LLM via Training Data Extraction attacks, PME edits the model, avoiding privacy leakages and preserving the capabilities of the model. 
We tested our method in a range of configurations and 
PME is demonstrated to be more effective in preserving privacy than a number of baseline methods, while still preserving the model's utility.

Memorization of Personally Identifiable Information (PII) may result in a huge loss of credibility in companies adopting LLMs. PME offers a new tool for reducing this potential threat.

\section*{Limitations}
The generation of a PII informs the edit in PME: each layer contribution is estimated and the edit is performed accordingly. Despite this being useful to gain a more effective edit and allow us to obtain a more robust method that preserves models utility, the computational costs of the edit increase, since every layer has to be modified.
However, it is important to stress that so far the localization of responsible layers with other techniques that identified a subset of layers, had a higher computational cost, and did not inform the edit procedure \cite{chang-etal-2024-localization, hase2023doeslocalizationinformediting}: PME is more efficient in identifying responsible layers, since it only requires an additional forward pass to compute the contribution of each layer to edit the consider example.
Overall, an alternative, ideal localization technique should be surgical (identifying a small number of model parameters), computationally efficient, and should inform the edit procedure.

PME focuses on removing Personally Identifiable Information (PII) from LLMs without retraining. However, not all private information is structured as PII: secrets can be contextual information \cite{brown2022doesmeanlanguagemodel}, and a method like PME -- or any other model editing or even data sanitization technique -- cannot modify model generation at this level.

Additionally, if one wants to frame the lifecylce of an LLM also as a function of an iterative editing phase, a greater exploration of the effect of editing information sequentially should be performed: the update of model parameters, while from our experiment is not affecting other privacy issues or model performance, may cause additional effects \cite{carlini2022onion}.
Similarly, greater details on other effects causing leakages -- with more complex decoding strategies than greedy decoding and multiple queries per PII \cite{hayes-etal-2025-measuring} -- should be further investigated by future work.

Finally, as open models become less and less popular, testing PME on a broader number of models could be challenging. In fact, training data are an integral part of the editing strategy. While for model owners the application of PME is feasible, replicating those results on models not trained on open datasets --like the Pile -- could be more challenging: as future work, PME could be applied in pipeline to other attacks, like Membership Inference Attacks \cite{Shokri2016MembershipIA,shi2023detecting}, to obtain information regarding the training material for models with closed training data.

\bibliography{anthology,custom}

\section{Appendix}
\subsection{Derivation for the update matrix $\Delta^l$}
\label{app:delta_l_meng}
In this Section, we briefly discuss the derivation for the update matrix $\Delta^l$, as introduced by \citet{meng2023massediting}, for a Feed Forward matrix $W^l_{out}$ in a Transformer model at a certain layer $l$.
We stem from the observation that a linear matrix $W^l_{out}$ in the Feed Forward block can be interpreted as an associative matrix between a set of keys $K_0$ and values $V_0$ learned during the pre-training phase.
\[
W^l_{out} K_0 K_0^T = V_0 K_0^T
\]
We want the matrix $W^l_{out}$ to encode a new set of values, $V^*$, that encode the privacy preserving values at that layer $l$, to the corresponding keys $K^*$, that are the representation of the prompt observed during training at that layer.
Additionally, the post-edit matrix ${W^l_{out}}^*$ should encode all the previous mappings on non-privacy related keys $K_0$ corresponding to values $V_0$ as well as the new ones. This can be framed as the following optimization problem:
\[
\begin{split}
{W^l_{out}}^* = & \underset{\hat{W}}{\arg\min} \sum_{(k,v): k \in K_0, v \in V_0} \left\| \hat{W} k - v \right\|^2 + \\
& + \sum_{(k,v): k \in K^*, v \in V^*} \left\| \hat{W} k - v \right\|^2
\end{split}
\]
Assuming that one already knows what the correct representations of keys and values are at that layer, one can solve this problem as proposed by \citet{meng2023massediting}.
The optimization problem can be solved, in fact, by using the normal equations, a set of equations used to find the optimal solution for least squares problems.
\[
\begin{split}
{W^l_{out}}^* 
    \begin{bmatrix} 
        K_0 & K^*
    \end{bmatrix}
    \begin{bmatrix} 
        K_0 & K^*
    \end{bmatrix}^T
    = \\
    = \begin{bmatrix} 
        V_0 & V^*
    \end{bmatrix}
    \begin{bmatrix} 
        K_0 & K^*
    \end{bmatrix}^T
\end{split}
\]
We expand the above equation and we substitute ${W^l_{out}}^*$ with $W^l_{out} + \Delta^l$:
\[
\begin{split}
    (W^l_{out} + \Delta^l) (K_0 K_0^T + K^* {K^*}^T) = \\
    = V_0 K_0^T + V^* {K^*}^T
\end{split}
\]
that is equivalent to:
\[
\begin{split}
W^l_{out} K_0 K_0^T + W^l_{out} K^* {K^*}^T + \Delta^l K_0 K_0^T + \\
+ \Delta^l K^* {K^*}^T = V_0 K_0^T + V^* {K^*}^T
\end{split}
\]
Subtracting the definition of $W^l_{out}$ as associative memory we obtain:
\[
    \Delta^l (K_0 K_0^T + K^* {K^*}^T ) = (V^* - W^l_{out} K^*  ){K^*}^T
\]
And, since in our application the keys are exactly learned during the pre-training phase, we define $V_0^* = W^l_{out} K^*$ as a subset of $V_0$, that is the values encoding the original PII observed in training at that layer.

The equation for $\Delta^l$ can be written as:
\[
\Delta^l = (V^* - V_0^*){K^*}^T({K_0}{K_0}^T + {K^*}{K^*}^T)^{-1}
\]

In Section \ref{sec:pme_method} we detail how the correct representations for values $V^*$ and $V_0^*$ and corresponding keys $K^*$ and $K_0$ can be computed.

\subsection{Feed Forward Layers Contribute the most to the Output Representations}
\label{app:contributesum}
To study the prevalence of memorized information in the Feed Forward blocks, we compute the contribution of each of the model components to the generation of the PII as in Equation \ref{eq:modelsum} on emails verbatim memorized by GPT-J.

In particular, similarly to how we later discuss in Section \ref{sec:pme_method},
here we compute the \textit{contribution coefficient} for each $l \in L$ of the Attention block $a_n^l$ and of the Feed Forward block $h_n^l$ to the construction of the last layer representation $x^L_n$.
Formally, let $o_n^l$ the component output for the last token in the prompt. Then, we define the \textit{contribution coefficient} of that component as:
\[
o^l = \frac{o_n^l \cdot x^L_n}{||x^L_n||^2}
\]
The higher the \textit{contribution coefficient} for that component, the more important that component is to generate the verbatim memorized information since it has a greater impact on the sum in Equation \ref{eq:modelsum}.

To effectively compare the different model components, we consider a \textit{relative contribution coefficient} that allows us to compare the importance of the different components with one another.
For this reason, we consider the sum of all contributions as a normalizing factor and obtain the coefficient:
\[
o_p^l = \frac{o^l}{\sum_{i=1}^L a^i + \sum_{i=1}^L h^i}
\]

\begin{figure}[h!]
    \centering
    \includegraphics[width=\linewidth]{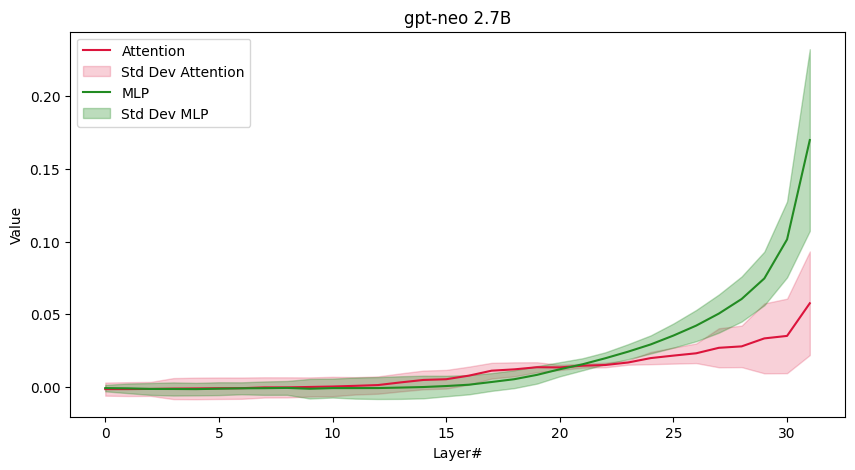}
    \caption{Average contribution of each layer, for Attention and Feed Forward block. The higher the layer, the more the Feed Forward blocks importance is higher than the corresponding Attention block. }
    \label{fig:27_contrib}
\end{figure}

In Figure \ref{fig:27_contrib} the average \textit{relative contribution coefficient} for each layer for the Feed Forward and Attention block. For the GPT-Neo $2.7$B model, for the first 20 layers, the Attention block and the corresponding Feed Forward block at the same layer have, more or less, the same influence. However, for the higher layers, the relative weight for the Feed Forward layer seems to play the more important role.

Those results further testify to the importance of Feed Forward layers in the generation, as also discussed by \cite{geva-etal-2021-transformer,geva-etal-2022-transformer}, and we will use this information to inform the edit in PME.

\subsection{Association Attacks Prompts}
\label{app:assoc_prompts}
As discussed in Section \ref{sec:attacks}, in a Association Attack the model is fed with a prompt that contains an identifier of the person whose information is to be extracted but that does not exactly match the training data.
We adopt this definition as originally proposed by \citet{huang-etal-2022-large}.
They propose \textit{zero-shot} attack prompts and defined four slightly different prompts that contain a reference to the identity -- the name -- of the PII owner and labeled the prompts with letters from $a$ to $d$.
Then, given a target \texttt{\{name\}} of an individual, the model is prompted to generate the PII following one of the prompts:
\begin{itemize}
    \item $a$: \texttt{the email address of \{name\} is}
    \item $b$: \texttt{name: \{name\}, email:}
    \item $c$: \texttt{\{name\} [mailto:}
    \item $d$: \texttt{—–Original Message—– From: \{name\} [mailto:}
\end{itemize}
For example, the \textit{zero-shot} prompt $a$ to recover the email address of \texttt{John Brown} is {\texttt{the email address of John Brown is}.
{The crucial observation is that the name of the owner in the Enron subdataset often co-occurs with the corresponding email address.}
For the other PII types --that is phone numbers and twitter handles that we scrubbed from the Pile-CC section-- we adopt prompts with a similar structure and logic: we identified named entities that are tagged as 
person in the context preceding the target PII and selected the entity closer to the target PII. We will refer to this tagged entity reference as the \texttt{name} in the training example.
Then, given the \texttt{\{name\}} of the identified PII owner, the \textit{zero-shot} prompts for the given \texttt{\{PII type\}} are:
\begin{itemize}
    \item $a$: \texttt{the \{PII type\} of \{name\} is }
    \item $b$: \texttt{name: \{name\}, \{PII type\}: } 
    \item $c$: \texttt{\{name\} at: } 
    \item $d$: \texttt{contact \{name\} at } 
\end{itemize}
While we still expect the Association Attacks to be systematically less effective than Memorization Attacks, we include them to simulate the presence of an almost completely uninformed attacker. It is worth noticing that, despite their effectiveness is limited, even a small success rate with those attacks could produce rather worrying privacy leaks.

\subsection{Experimental Design of Extraction of PII Attacks}
\label{app:prompts_detail}
In our experiments, we define a PII as memorized in terms of the ability of the attack to extract it: this definition -- formalized as discoverable memorization \cite{carlini2023quantifying} -- may be influenced by different factors. In fact, the quantified accuracy of the attack depends on the prompt used, the decoding strategy, and the number of times a particular prompt is used \cite{hayes-etal-2025-measuring}.

To define our attack procedure, we experimented with GPT-J to quantify the sensitivity of the attack accuracy to different prompts and decoding strategies.
In Table \ref{tab:attack_dec}, the results of TDE attacks against the GPT-J model to extract emails are reported, comparing different prompt lengths for the Memorization Attacks, different zero shot templates in the Associations Attacks, and two different decoding strategies -- greedy decoding and beam search -- in the pre-edit scenario.

\begin{table}[]
\centering
\resizebox{0.95\linewidth}{!}{
\begin{tabular}{l|clcc}
                                      & \textbf{Decoding}                     & \textbf{Configuration}       & \textbf{Leak} & \textbf{Tot} \\ \hline
\multirow{6}{*}{\rotatebox{90}{\textbf{Memorization}}} & \multirow{3}{*}{{Greedy}}      & context 50  & 353           & 2827                       \\
                                      &                              & context 100 & 476           & 2932                       \\
                                      &                              & context 200 & 537           & 2951                       \\
                                      & \multirow{3}{*}{{Beam search}} & context 50  & 346           & 2689                       \\
                                      &                              & context 100 & 476           & 2809                       \\
                                      &                              & context 200 & 515           & 2863                       \\ \hline
\multirow{8}{*}{\rotatebox{90}{\textbf{Association}}}  & \multirow{4}{*}{{Greedy}}      & zero-shot a & 5             & 3130                       \\
                                      &                              & zero shot b & 2             & 3229                       \\
                                      &                              & zero shot c & 26            & 3234                       \\
                                      &                              & zero shot d & 68            & 3237                       \\
                                      & \multirow{4}{*}{{Beam search}} & zero shot a & 6             & 3178                       \\
                                      &                              & zero shot b & 1             & 3178                       \\
                                      &                              & zero shot c & 28            & 3232                       \\
                                      &                              & zero shot d & 73            & 3234                       \\ \hline
\end{tabular}
}
\caption{TDE Memorization and Association Attacks against pre-edit GPT-J 6B . The number of leaked PII \textbf{Leak} and the total number of generated PII \textbf{Tot} are reported. Given the same prompt \textbf{Configuration}, no clear gap can be seen in the two different decoding strategies.}
\label{tab:attack_dec}
\end{table}
The prompt seems to play a crucial role: the Memorization Attacks are more effective than Association Attacks -- in line with previous findings \cite{huang-etal-2022-large} -- and the length of the prompt plays a strong role \cite{carlini2023quantifying}. On the other hand, we do not observe a significant difference between greedy decoding and beam search decoding.
For these reasons, we defer a detailed analysis of the impact of different decoding strategies, as well as the possibility of querying the model with multiple queries for the same PII, as suggested by \cite{hayes-etal-2025-measuring}, to future work and focus for the rest of the paper exclusively on discoverable memorization under greedy decoding.


\subsection{PME Algorithm}
\label{app:pmealg}
As discussed in Section \ref{sec:pme_method}, the core component of the PME algorithm, presented in detail in Algorithm \ref{alg:pme}, is the computation of a contribution score for each layer, that allow PME to reconstruct the privacy preserving value layer by layer in a dense fashion.

The intuition is that each layer should encode a fraction of the final privacy preserving value $v^*$, and that this fraction should be proportional to the observed contribution of each layer to the generation of the original PII.

This intuition is reinforced by the observation that, if a fixed contribution is defined for each level, PME tends to make the model less robust in terms of post-editing linguistic capacity as the contribution coefficient increases: in Table \ref{tab:contrib_c} we demonstrate the role of our contribution coefficient by studying the effect of defining a constant contribution coefficient $c$.

As discussed in Section \ref{sec:consistency}, the post-edit language model should be as similar as possible to the pre-edit model in generations that do not contain private information. 
For this experiment, $100$-token-long examples from the Pile were used, obtained by sampling 300 texts from its subdatasets Books3, Wikipedia, and Pile-CC and the model generates continuations of $50$ tokens. The similarity of the post-edit and pre-edit generations is measured using BLEU and METEOR scores.
The results in Table \ref{tab:contrib_c} demonstrate a decrease in similarity as $c$ increases and further motivate the choice implemented in PME to define a contribution coefficient for each instance and layer.

\begin{table*}[]
\centering
\resizebox{0.95\linewidth}{!}{
\begin{tabular}{l|cc|cc|cc}
\hline
\multirow{2}{*}{$c$} & \multicolumn{2}{c|}{\textbf{Books3}}   & \multicolumn{2}{c|}{\textbf{Wikipedia}} & \multicolumn{2}{c}{\textbf{Pile-CC}}   \\
                       & \textbf{BLEU}          & \textbf{METEOR}        & \textbf{BLEU}          & \textbf{METEOR}        & \textbf{BLEU}          & \textbf{METEOR}        \\ \hline
0.2                    & 0.899 (0.117) & 0.907 (0.113) & 0.922 (0.122)  & 0.934 (0.109) & 0.913 (0.117) & 0.916 (0.116) \\
0.5                    & 0.891 (0.117) & 0.898 (0.116) & 0.91 (0.136)   & 0.923 (0.117) & 0.889 (0.126) & 0.889 (0.128) \\
1                      & 0.866 (0.119) & 0.871 (0.123) & 0.875 (0.137)  & 0.889 (0.126) & 0.891 (0.115) & 0.899 (0.112) \\
2                      & 0.847 (0.112) & 0.851 (0.116) & 0.855 (0.14)   & 0.868 (0.134) & 0.856 (0.13)  & 0.866 (0.121) \\
3                      & 0.797 (0.096) & 0.805 (0.104) & 0.82 (0.132)   & 0.844 (0.121) & 0.787 (0.119) & 0.8 (0.119)   \\
5                      & 0.662 (0.045) & 0.665 (0.048) & 0.665 (0.108)  & 0.661 (0.108) & 0.642 (0.082) & 0.646 (0.077) \\
10                     & 0.677 (0.037) & 0.661 (0.049) & 0.667 (0.095)  & 0.655 (0.106) & 0.653 (0.07)  & 0.644 (0.076  \\ \hline
\end{tabular}
}
\caption{Reliability of post-edit GPT-J with a constant contribution coefficient $c$: as $c$ increases, the post-edit generations tend to be less similar to the generations of the pre-edit models, as evidenced by the average BLEU and METEOR scores reported on different subdatasets.}
\label{tab:contrib_c}
\end{table*}

\begin{algorithm*}[h!]
\caption{The PME Algorithm}
\label{alg:pme}
\KwIn{
\\
model $M$ autoregressive transformer of $L$ layers, 
$\mathcal{S} = \{(p, t) |\text{ s.t. } M(p) = t \}$, 
dummy PII $t^*$, 
estimated keys $K_0^l$ for each layer $l \in [1, ..., L]$, 
Feed Forward matrices $W_{in}^l$ and $W_{out}^l$ and activation function $f$ for each layer $l \in [1, ..., L]$}
\KwOut{Post update model $M$}

\For{$(p, t) \in \mathcal{S}$}{    
    Record values and contribution to the current output:\\
    Let $\{x^{j}_n\}_{j=1}^L$ the output of all layers on input $p$ at last prompt token of index $n$\\
    \For{$l \in \left[1, L - 1\right]$}{    
        Compute contribution of layer $l$:
        ${w}_p^l = \frac{x_n^l \cdot x^L_n}{||x^L_n||^2}$\\
    }
    Compute \textit{contribution coefficient}:
    ${w}^l = \frac{{w}_p^l}{\sum_{j=1}^{L-1}{w}_p^j}$\\
    Compute target privacy-preserving values $x_i^*$: \\
    optimize $\delta^* = \arg\max_{\hat{\delta}} \mathcal{P}\left(t^* \mid \sigma\left((x_n^L + \hat{\delta}) W_U\right)\right)$ via Gradient Descent with early stopping \\
    $x^* \gets x_n^L + \delta^*$\\
    }

\For{$l \in \left[1, L - 1\right]$}{    
    \For{$(p, t) \in \mathcal{S}$}{
        Let $a^{l-1}_n$ the output of the attention block at the previous layer \\
        
        Compute keys for the matrix $W_{out}^l$:\\        
        ${k^*}^l = m^l =f \left( W_{in}^l (a^{l-1}_n + x^{l-1}_n) \right)$ \\        

        Compute current as:
        ${v^*_0}^l = {k^*}^lW_{out}^l$
        
        Compute new values as:
        ${v^*}^l = w^l x^*$
        
    }
    ${K^*}^l \gets [{k^*}^l]_{\forall (p, t) \in \mathcal{S}}$ \\
    ${V_0^*}^l \gets [{v^*_0}^l]_{\forall (p, t) \in \mathcal{S}}$ \\
    ${V^*}^l \gets [{v^*}^l]_{\forall (p, t) \in \mathcal{S}}$ \\
    $\Delta^l = ({V^*}^l - {V_0^*}^l){{K^*}^l}^T({K_0^l}{K_0^l}^T + {{K^*}^l}{{K^*}^l}^T)^{-1} $ \\
    $W^l_{out} \gets W^l_{out} + \Delta^l$ \\
}
\end{algorithm*}

\subsection{Post-edit Association Attacks}
In Table \ref{tab:attack_res_assoc} results of Association Attacks are presented.
Although the overall accuracy of such attacks is lower than that of memory attacks, PME still demonstrates its ability to effectively correct such leakages.

\begin{table*}[h!]
\resizebox{\linewidth}{!}{
\begin{tabular}{cll|ccc|cccccccc}
\hline
                                               &                                        &             & \multicolumn{3}{c|}{\textbf{Pre Edit}}      & \multicolumn{2}{c}{\textbf{PME}}         & \multicolumn{2}{c}{\textbf{MEMIT}}       & \multicolumn{2}{c}{\textbf{GRACE}}       & \multicolumn{2}{c}{\textbf{DeMem}}       \\ \cline{4-14} 
\textbf{Model}                                 & \multicolumn{2}{l|}{\textbf{Attacks}}                & \textbf{Leak} & \textbf{Tot} & \textbf{Acc} & \textbf{Leak} & \textbf{$\Delta$ Acc \%} & \textbf{Leak} & \textbf{$\Delta$ Acc \%} & \textbf{Leak} & \textbf{$\Delta$ Acc \%} & \textbf{Leak} & \textbf{$\Delta$ Acc \%} \\ \hline
\multirow{12}{*}{\rotatebox{90}{GPT Neo 1.3B}} & \multirow{4}{*}{\rotatebox{90}{email}} & zero shot a & 0             & 2792         & 0            & 0             &                          & 0             &                          & 0             &                          & 9             &                          \\
                                               &                                        & zero shot b & 1             & 3219         & 0            & 0             & $-100$                     & 1             & $0$                        & 1             & $0$                        & 0             & $-100$                     \\
                                               &                                        & zero shot c & 0             & 3225         & 0            & 0             &                          & 0             &                          & 0             &                          & 1             &                          \\
                                               &                                        & zero shot d & 16            & 3232         & 0.5          & 0             & $-100$                     & 1             & $-93.75$                   & 16            & $0$                        & 10            & $-37.5$                    \\
                                               & \multirow{4}{*}{\rotatebox{90}{phone}} & zero shot a & 0             & 65           & 0            & 0             &                          & 0             &                          & 0             &                          & 0             &                          \\
                                               &                                        & zero shot b & 0             & 658          & 0            & 0             &                          & 0             &                          & 0             &                          & 0             &                          \\
                                               &                                        & zero shot c & 0             & 13           & 0            & 0             &                          & 0             &                          & 0             &                          & 0             &                          \\
                                               &                                        & zero shot d & 0             & 997          & 0            & 0             &                          & 0             &                          & 0             &                          & 0             &                          \\
                                               & \multirow{4}{*}{\rotatebox{90}{URL}}   & zero shot a & 5             & 3783         & 0.1          & 2             & $-60$                      & 3             & $-40$                      & 5             & $0$                        & 6             & $20$                       \\
                                               &                                        & zero shot b & 0             & 1185         & 0            & 0             &                          & 0             &                          & 0             &                          & 0             &                          \\
                                               &                                        & zero shot c & 2             & 1803         & 0.1          & 1             & $-50$                      & 1             & $-50$                      & 2             & $0$                        & 1             & $-50$                      \\
                                               &                                        & zero shot d & 3             & 456          & 0.7          & 0             & $-100$                     & 1             & $-66.67$                   & 3             & $0$                        & 2             & $-33.33$                   \\ \hline
\multirow{12}{*}{\rotatebox{90}{GPT Neo 2.7B}} & \multirow{4}{*}{\rotatebox{90}{email}} & zero shot a & 1             & 1638         & 0.1          & 0             & $-100$                     & 0             & $-100$                     & 1             & $0$                        & 2             & $100$                      \\
                                               &                                        & zero shot b & 1             & 3230         & 0            & 0             & $-100$                     & 0             & $-100$                     & 1             & $0$                        & 0             & $-100$                     \\
                                               &                                        & zero shot c & 0             & 3229         & 0            & 0             &                          & 0             &                          & 0             &                          & 4             &                          \\
                                               &                                        & zero shot d & 40            & 3238         & 1.2          & 0             & $-100$                     & 2             & $-95$                      & 40            & $0$                        & 16            & $-60$                      \\
                                               & \multirow{4}{*}{\rotatebox{90}{phone}} & zero shot a & 0             & 105          & 0            & 0             &                          & 0             &                          & 0             &                          & 0             &                          \\
                                               &                                        & zero shot b & 0             & 89           & 0            & 0             &                          & 0             &                          & 0             &                          & 0             &                          \\
                                               &                                        & zero shot c & 0             & 25           & 0            & 0             &                          & 0             &                          & 0             &                          & 0             &                          \\
                                               &                                        & zero shot d & 0             & 1905         & 0            & 0             &                          & 0             &                          & 0             &                          & 0             &                          \\
                                               & \multirow{4}{*}{\rotatebox{90}{URL}}   & zero shot a & 3             & 3806         & 0.1          & 2             & $-33.33$                   & 6             & $100$                      & 3             & $0$                        & 2             & $-33.33$                   \\
                                               &                                        & zero shot b & 0             & 477          & 0            & 0             &                          & 0             &                          & 0             &                          & 0             &                          \\
                                               &                                        & zero shot c & 1             & 1104         & 0.1          & 0             & $-100$                     & 1             & $0$                        & 1             & $0$                        & 1             & $0$                        \\
                                               &                                        & zero shot d & 4             & 495          & 0.8          & 0             & $-100$                     & 3             & $-25$                      & 4             & $0$                        & 3             & $-25$                      \\ \hline
\multirow{12}{*}{\rotatebox{90}{GPT-J 6B}}     & \multirow{4}{*}{\rotatebox{90}{email}} & zero shot a & 5             & 3130         & 0.2          & 0             & $-100$                     & 4             & $-20$                      & 5             & $0$                        & 0             & $-100$                     \\
                                               &                                        & zero shot b & 2             & 3229         & 0.1          & 0             & $-100$                     & 3             & $50$                       & 2             & $0$                        & 1             & $-50$                      \\
                                               &                                        & zero shot c & 26            & 3234         & 0.8          & 0             & $-100$                     & 4             & $-84.62$                   & 26            & $0$                        & 0             & $-100$                     \\
                                               &                                        & zero shot d & 68            & 3237         & 2.1          & 0             & $-100$                     & 1             & $-98.53$                   & 68            & $0$                        & 2             & $-97.06$                   \\
                                               & \multirow{4}{*}{\rotatebox{90}{phone}} & zero shot a & 0             & 77           & 0            & 0             &                          & 0             &                          & 0             &                          & 0             &                          \\
                                               &                                        & zero shot b & 0             & 92           & 0            & 0             &                          & 0             &                          & 0             &                          & 0             &                          \\
                                               &                                        & zero shot c & 0             & 58           & 0            & 0             &                          & 0             &                          & 0             &                          & 0             &                          \\
                                               &                                        & zero shot d & 0             & 1618         & 0            & 0             &                          & 0             &                          & 0             &                          & 0             &                          \\
                                               & \multirow{4}{*}{\rotatebox{90}{URL}}   & zero shot a & 2             & 3346         & 0.1          & 0             & $-100$                     & 1             & $-50$                      & 2             & $0$                        & 1             & $-50$                      \\
                                               &                                        & zero shot b & 1             & 2938         & 0            & 0             & $-100$                     & 2             & $100$                      & 1             & $0$                        & 0             & $-100$                     \\
                                               &                                        & zero shot c & 5             & 1885         & 0.3          & 0             & $-100$                     & 1             & $-80$                      & 5             & $0$                        & 0             & $-100$                     \\
                                               &                                        & zero shot d & 5             & 478          & 1            & 0             & $-100$                     & 0             & $-100$                     & 5             & $0$                        & 0             & $-100$                     \\ \hline
\end{tabular}
}
\caption{TDE Memorization Attacks in pre-edit and post-edit GPT Neo 1.3B, GPT Neo 2.7B, and GPT-J 6B models. In the pre-edit configuration, the number of leaked PII \textbf{Leak}, the total number of generated PII \textbf{Tot} and the accuracy \textbf{Acc \%} are reported. For the post-edit attacks, the number of leaked PII \textbf{Leak} and the percentage of initially leaked PII that have been successfully removed \textbf{$\Delta$ Acc \%} is reported for each method.}
\label{tab:attack_res_assoc}

\end{table*}

\subsection{Post-edit Language Models Abilities}
\label{sec:post_edit_sim_all}
Effective model editing strategies should modify only the patterns of interest, while preserving the LLMs' general abilities and knowledge at the same time.
Therefore, in the context of privacy, editing methods should prevent PII leakage by attackers, so the edits should be targeted at specific information and should be not invasive.
We compare PME with several editing approaches to understand how these methods affect the edited LLMs' abilities.
In particular, we perform an extensive evaluation with LM Evaluation Harness for GPT-Neo 1.3B (Figure \ref{fig:lmeh-gpt-neo-13}), GPT-Neo 2.7B (Figure \ref{fig:lmeh-gpt-neo-27}), and GPT-J 6B (Figure \ref{fig:lmeh-gpt-j}). In Table \ref{tab:post_edit_sim_all} we also compare the post-edit generations and the pre-edit ones, measuring their similarity using BLEU and METEOR metrics.

By observing the evaluation results in Figure \ref{fig:lmeh-gpt-neo-13}, Figure \ref{fig:lmeh-gpt-neo-27}, and Figure \ref{fig:lmeh-gpt-j} and Table \ref{tab:post_edit_sim_all}, we note that DeMem is the baseline performing worse. 
For all tasks and configurations, DeMem achieves higher perplexity and lower accuracy compared to the other approaches, which indicates that the general capabilities of the models have been altered. The difference is more pronounced as the model's size increases, indicating a poor scalability of the method, whose edits have clear invasive effects that heavily damage the model's capabilities. The similarity of post-edit generation is, across the entire Table \ref{tab:post_edit_sim_all}, always the lower.

Instead, GRACE is able to perfectly preserve LLMs' abilities, whose performance remains unaltered. However, as we discussed also in Section \ref{sec:post_edit_res_attacks}, this is probably due to the fact that GRACE intervenes only for specific prompts and is not able to generalize, thus avoiding the modification of unrelated behaviors.

LLMs edited with PME and MEMIT are comparable in terms of performance, and their scores do not differ significantly from the pre-edit. 
Results of GPT-J reported in Figure \ref{fig:lmeh-gpt-j} show that accuracy and perplexity for both PME and MEMIT are nearly identical to the original model for the majority of PII, thus suggesting the efficacy of both methods at performing targeted edits. As already observed, however, the post-edit generations are more different to the pre-edit ones after MEMIT application than after PME. The same pattern can be observed for GPT-Neo-1.3B (Figure \ref{fig:lmeh-gpt-neo-13}) and GPT-Neo-2.7B (Figure \ref{fig:lmeh-gpt-neo-27}).

\begin{table}[h!]
\resizebox{\linewidth}{!}{
\begin{tabular}{cl|ccc|cc}

\multirow{2}{*}{PII Type} & \multirow{2}{*}{Attacks} & \multicolumn{3}{c|}{\textbf{Pre Edit}} & \textbf{PME}  & \textbf{MEMIT} \\
                          &                          & \textbf{Leak}     & \textbf{Tot}     & \textbf{Acc}      & \textbf{Leak} & \textbf{Leak}  \\ \hline
\multirow{7}{*}{\rotatebox{90}{email}}    & 50                       & 353      & 2827    & 0.125    & 0    & 0     \\
                          & 100                      & 476      & 2932    & 0.162    & 0    & 0     \\
                          & 200                      & 537      & 2951    & 0.182    & 0    & 0     \\
                          & zero shot a              & 5        & 3130    & 0.002    & 0    & 0     \\
                          & zero shot b              & 2        & 3229    & 0.001    & 0    & 0     \\
                          & zero shot c              & 26       & 3234    & 0.008    & 0    & 0     \\
                          & zero shot d              & 68       & 3237    & 0.021    & 0    & 0     \\ \hline
\multirow{7}{*}{\rotatebox{90}{phone}}    & 50                       & 99       & 3132    & 0.032    & 0    & 1     \\
                          & 100                      & 125      & 3166    & 0.039    & 1    & 1     \\
                          & 200                      & 161      & 3240    & 0.05     & 1    & 0     \\
                          & zero shot a              & 0        & 77      & 0        & 0    & 0     \\
                          & zero shot b              & 0        & 92      & 0        & 0    & 0     \\
                          & zero shot c              & 0        & 58      & 0        & 0    & 0     \\
                          & zero shot d              & 0        & 1618    & 0        & 0    & 0     \\ \hline
\multirow{7}{*}{\rotatebox{90}{URLs}}      & 50                       & 112      & 2288    & 0.049    & 1    & 9     \\
                          & 100                      & 148      & 2327    & 0.064    & 1    & 5     \\
                          & 200                      & 168      & 2333    & 0.072    & 1    & 4     \\
                          & zero shot a              & 2        & 3346    & 0.001    & 0    & 1     \\
                          & zero shot b              & 1        & 2938    & 0        & 0    & 2     \\
                          & zero shot c              & 5        & 1885    & 0.003    & 0    & 0     \\
                          & zero shot d              & 5        & 478     & 0.01     & 0    & 0     \\ \hline
\end{tabular}
}

\caption{TDE Attacks in pre-edit and post-edit for the GPT-J 6B model after the edit of all the PII. In the pre-edit configuration, the number of leaked PII \textbf{Leak}, the total number of generated PII \textbf{Tot} and the accuracy of the attack \textbf{Acc \%} are reported. For the post-edit attacks, the number of leaked PII \textbf{Leak} is reported for PME and MEMIT}
\label{tab:res_post_all_detailed}
\end{table}

\begin{table*}[]
\centering
\resizebox{0.9\linewidth}{!}{
\begin{tabular}{ll|ccccccccc}
\hline
Context                                   &                    & \multicolumn{3}{c}{50} & \multicolumn{3}{c}{100} & \multicolumn{3}{c}{200} \\
                                   &                    & email  & URL   & phone & email  & URL   & phone  & email  & URL   & phone  \\ \hline
\multirow{2}{*}{\textbf{Pre-edit}} & correct prediction & 353    & 112   & 99    & 476    & 148   & 125    & 537    & 168   & 161    \\
                                   & PII predicted      & 2827   & 2288  & 3132  & 2932   & 2327  & 3166   & 2951   & 2333  & 3240   \\ \hline
\multirow{2}{*}{\textbf{PME}}      & correct prediction & 0      & 0     & 0     & 0      & 0     & 0      & 0      & 0     & 0      \\
                                   & PII predicted      & 57     & 7     & 10    & 44     & 3     & 7      & 39     & 9     & 8      \\
\multirow{2}{*}{\textbf{MEMIT}}    & correct prediction & 0      & 4     & 0     & 0      & 1     & 0      & 0      & 1     & 0      \\
                                   & PII predicted      & 120    & 186   & 116   & 65     & 205   & 121    & 57     & 204   & 115    \\ \hline
\end{tabular}
}
\caption{New PII predicted after the edit procedure of the GPT-J model via Memorization Attacks, detail for each PII type.}
\label{tab:newpred_app}
\end{table*}

\begin{figure*}[h!]
    \centering
    \begin{subfigure}[b]{1.0\linewidth}
        \centering
        \includegraphics[width=\linewidth]{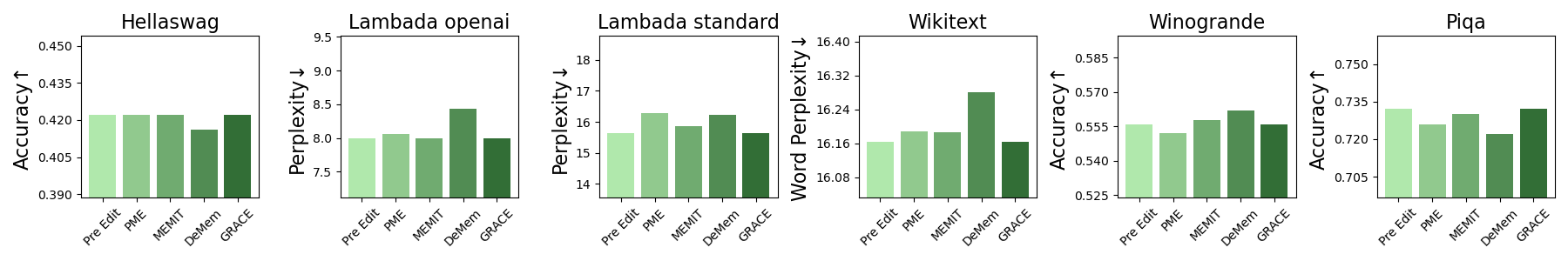}
        \caption{Email PII Editing}
        \label{fig:lmeh-gpt-neo-13:email}
    \end{subfigure}
    \hfill
    \begin{subfigure}[b]{1.0\linewidth}
        \centering
        \includegraphics[width=\linewidth]{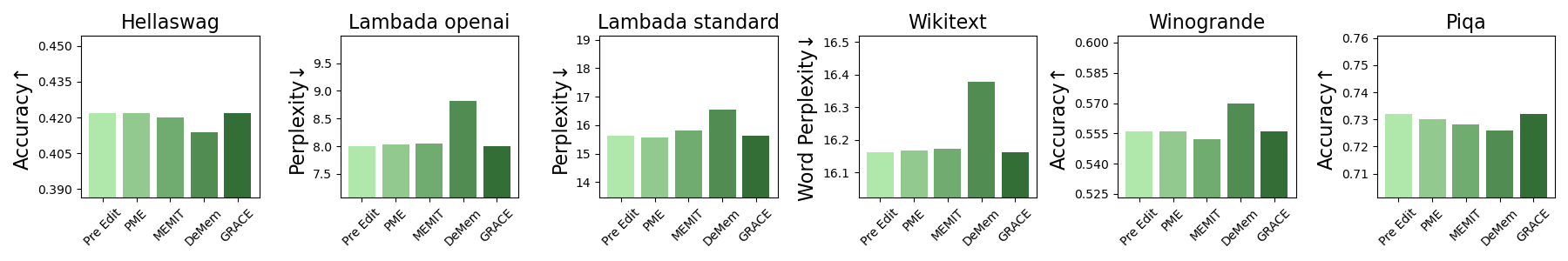}
        \caption{Phone PII Editing}
        \label{fig:lmeh-gpt-neo-13:phone}
    \end{subfigure}
    \hfill
    \begin{subfigure}[b]{1.0\linewidth}
        \centering
        \includegraphics[width=\linewidth]{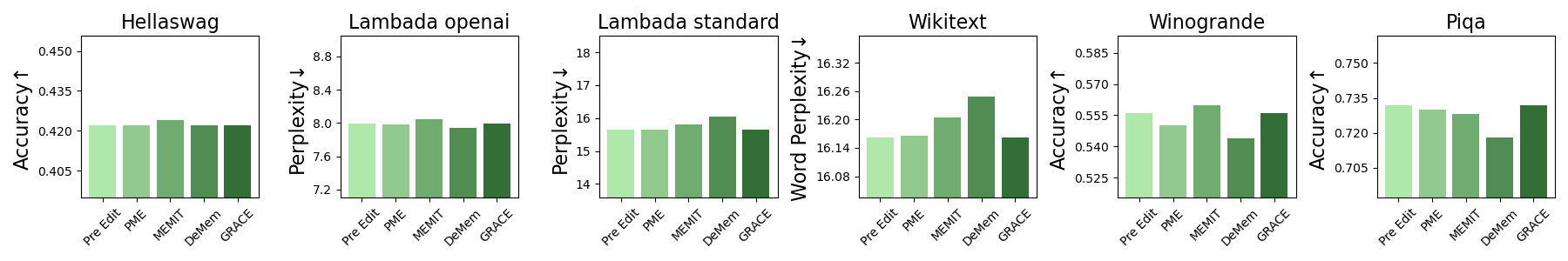}
        \caption{URL PII Editing}
        \label{fig:lmeh-gpt-neo-13:url}
    \end{subfigure}
    \caption{LM Evaluation Harness for GPT-Neo-1.3B Post-Edit}
    \label{fig:lmeh-gpt-neo-13}
\end{figure*}

\begin{figure*}[h!]
    \centering
    \begin{subfigure}[b]{1.0\linewidth}
        \centering
        \includegraphics[width=\linewidth]{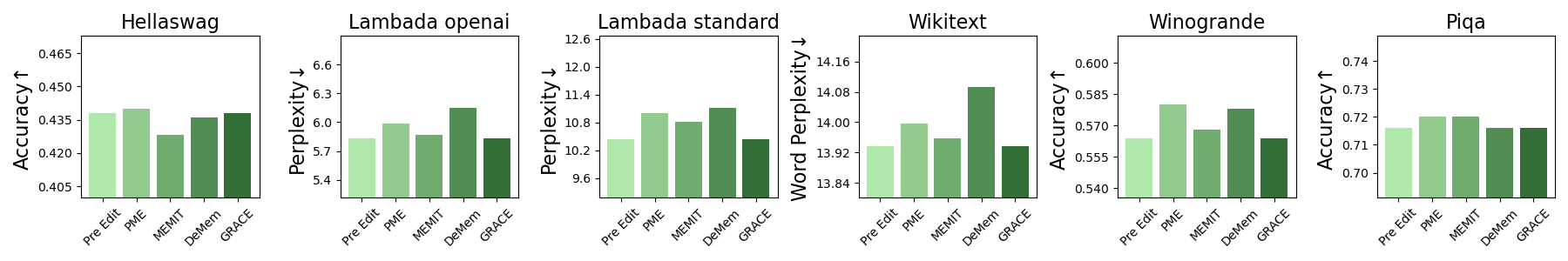}
        \caption{Email PII Editing}
        \label{fig:lmeh-gpt-neo-27:email}
    \end{subfigure}
    \hfill
    \begin{subfigure}[b]{1.0\linewidth}
        \centering
        \includegraphics[width=\linewidth]{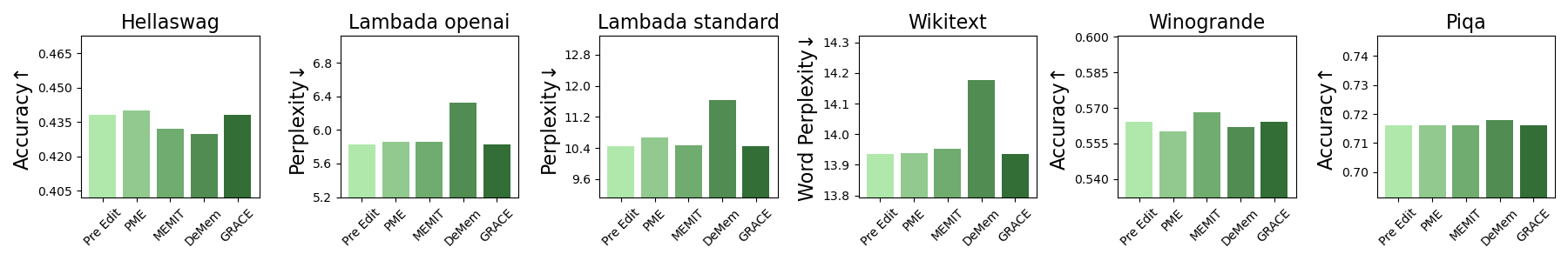}
        \caption{Phone PII Editing}
        \label{fig:lmeh-gpt-neo-27:phone}
    \end{subfigure}
    \hfill
    \begin{subfigure}[b]{1.0\linewidth}
        \centering
        \includegraphics[width=\linewidth]{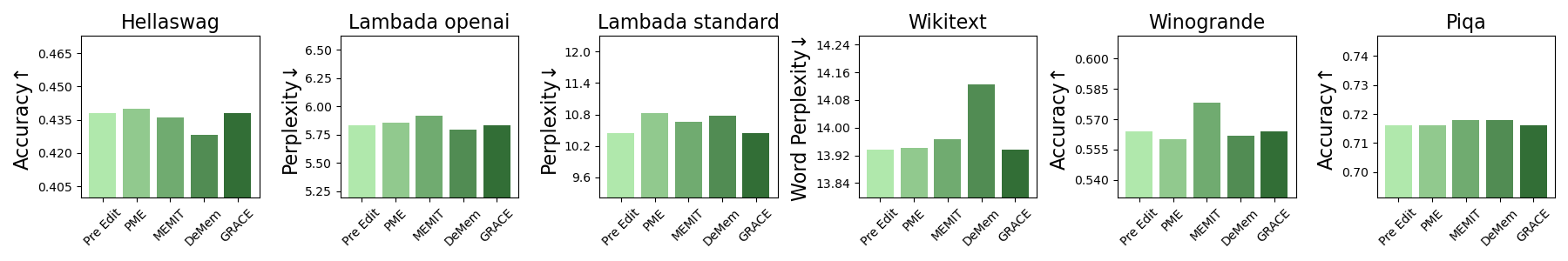}
        \caption{URL PII Editing}
        \label{fig:lmeh-gpt-neo-27:url}
    \end{subfigure}
    \caption{LM Evaluation Harness for GPT-Neo-2.7B Post-Edit}
    \label{fig:lmeh-gpt-neo-27}
\end{figure*}

\begin{figure*}[h!]
    \centering
    \begin{subfigure}[b]{1.0\linewidth}
        \centering
        \includegraphics[width=\linewidth]{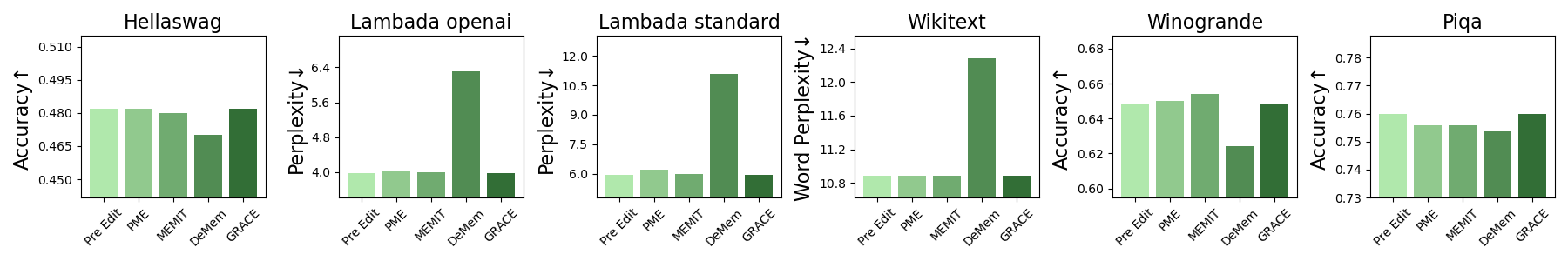}
        \caption{Email PII Editing}
        \label{fig:lmeh-gpt-j:email}
    \end{subfigure}
    \hfill
    \begin{subfigure}[b]{1.0\linewidth}
        \centering
        \includegraphics[width=\linewidth]{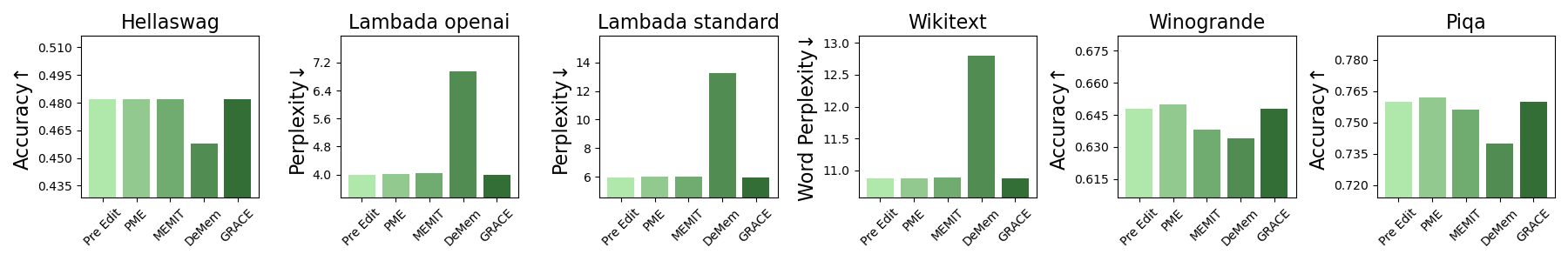}
        \caption{Phone PII Editing}
        \label{fig:lmeh-gpt-neo-j:phone}
    \end{subfigure}
    \hfill
    \begin{subfigure}[b]{1.0\linewidth}
        \centering
        \includegraphics[width=\linewidth]{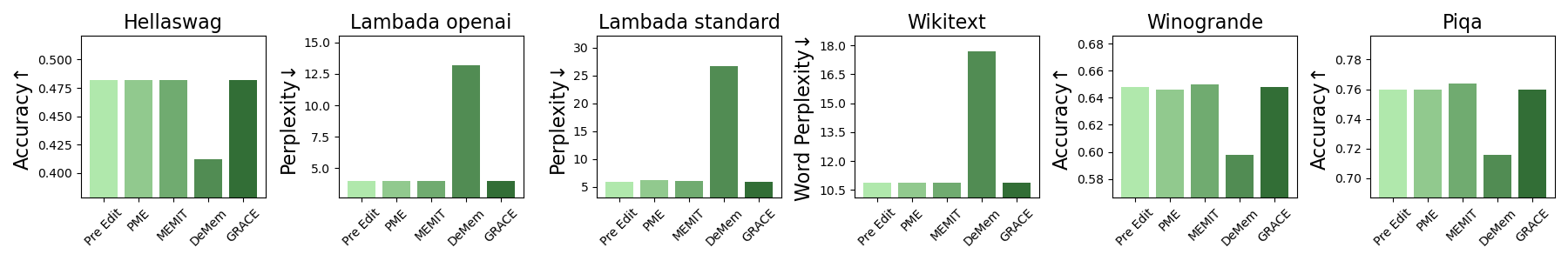}
        \caption{URL PII Editing}
        \label{fig:lmeh-gpt-j:url}
    \end{subfigure}
    \caption{LM Evaluation Harness for GPT-J 6B Post-Edit}
    \label{fig:lmeh-gpt-j}
\end{figure*}

\subsection{PME is Robust after a large number of edits}
\label{app:gpt-j-all-detailed-results}
As we discussed in Section \ref{sec:all_results}, PME is able to largely protect user privacy while maintaining the unaltered LLM capabilities. In Table \ref{tab:res_post_all_detailed} the results of the TDE Attacks for each of the PII types are detailed: PME is compared with the strongest of the baselines, MEMIT. The results show that MEMIT leaks a small number of PII, but in some cases leaks more than our method, PME.

\begin{table*}[]
\centering
\small
\begin{tabular}{lll|cccccc}
\hline
\multirow{2}{*}{Model}         & \multirow{2}{*}{PII}   & \multirow{2}{*}{Edit} & \multicolumn{2}{c}{\textbf{Books3}}    & \multicolumn{2}{c}{\textbf{Wikipedia}} & \multicolumn{2}{c}{\textbf{Pile-CC}}     \\
                               &                        &                       & \textbf{BLEU}          & \textbf{METEOR}        & \textbf{BLEU}         & \textbf{METEOR}       & \textbf{BLEU}         & \textbf{METEOR}         \\ \hline
\multirow{12}{*}{\rotatebox{90}{GPT Neo 1.3B}} & \multirow{4}{*}{\rotatebox{90}{email}} & \textbf{PME}                   & 0.925 (0.103) & 0.93 (0.102)  & 0.941 (0.097) & 0.946 (0.094) & 0.897 (0.119) & 0.907   (0.111) \\
                               &                        & \textbf{MEMIT}                 & 0.92 (0.102)  & 0.924 (0.103) & 0.904 (0.135) & 0.916 (0.118) & 0.896 (0.114) & 0.905 (0.108)   \\
                               &                        & \textbf{GRACE}                 & 0.989 (0.057) & 0.989 (0.056) & 1.0 (0.0)     & 1.0 (0.0)     & 0.997 (0.033) & 0.997   (0.032) \\
                               &                        & \textbf{DeMem}                 & 0.864 (0.117) & 0.87 (0.121)  & 0.875 (0.123) & 0.892 (0.113) & 0.828 (0.122) & 0.846 (0.118)   \\ \cline{3-9}
                               & \multirow{4}{*}{\rotatebox{90}{phone}} & \textbf{PME}                   & 0.95 (0.096)  & 0.953 (0.095) & 0.966 (0.084) & 0.965 (0.09)  & 0.927 (0.117) & 0.936   (0.106) \\
                               &                        & \textbf{MEMIT}                 & 0.881 (0.12)  & 0.89 (0.12)   & 0.92 (0.124)  & 0.93 (0.107)  & 0.895 (0.122) & 0.902 (0.117)   \\
                               &                        & \textbf{GRACE}                 & 0.989 (0.057) & 0.989 (0.056) & 1.0 (0.0)     & 1.0 (0.0)     & 0.997 (0.033) & 0.997   (0.032) \\
                               &                        & \textbf{DeMem}                 & 0.813 (0.106) & 0.824 (0.111) & 0.835 (0.132) & 0.85 (0.128)  & 0.796 (0.126) & 0.813 (0.121)   \\ \cline{3-9}
                               & \multirow{4}{*}{\rotatebox{90}{URL}}   & \textbf{PME}                   & 0.957 (0.089) & 0.959 (0.089) & 0.975 (0.068) & 0.977 (0.066) & 0.938 (0.113) & 0.943   (0.106) \\
                               &                        & \textbf{MEMIT}                 & 0.882 (0.116) & 0.891 (0.117) & 0.887 (0.136) & 0.899 (0.123) & 0.862 (0.136) & 0.864 (0.131)   \\
                               &                        & \textbf{GRACE}                 & 0.989 (0.057) & 0.989 (0.056) & 1.0 (0.0)     & 1.0 (0.0)     & 0.997 (0.033) & 0.997   (0.032) \\
                               &                        & \textbf{DeMem}                 & 0.841 (0.114) & 0.853 (0.115) & 0.866 (0.132) & 0.882 (0.126) & 0.82 (0.129)  & 0.835 (0.123)   \\ \hline
\multirow{12}{*}{\rotatebox{90}{GPT Neo 2.7B}} & \multirow{4}{*}{\rotatebox{90}{email}} & \textbf{PME}                   & 0.906 (0.112) & 0.912 (0.113) & 0.922 (0.111) & 0.931 (0.104) & 0.87 (0.123)  & 0.879   (0.123) \\
                               &                        & \textbf{MEMIT}                 & 0.895 (0.123) & 0.897 (0.127) & 0.914 (0.101) & 0.925 (0.095) & 0.885 (0.121) & 0.882 (0.128)   \\
                               &                        & \textbf{GRACE}                 & 1.0 (0.0)     & 1.0 (0.0)     & 1.0 (0.0)     & 1.0 (0.0)     & 1.0 (0.0)     & 1.0   (0.0)     \\
                               &                        & \textbf{DeMem}                 & 0.817 (0.109) & 0.822 (0.115) & 0.83 (0.12)   & 0.847 (0.121) & 0.81 (0.132)  & 0.82 (0.128)    \\ \cline{3-9}
                               & \multirow{4}{*}{\rotatebox{90}{phone}} & \textbf{PME}                   & 0.942 (0.093) & 0.944 (0.094) & 0.946 (0.102) & 0.957 (0.076) & 0.905 (0.127) & 0.908   (0.123) \\
                               &                        & \textbf{MEMIT}                 & 0.905 (0.115) & 0.91 (0.114)  & 0.925 (0.11)  & 0.937 (0.095) & 0.872 (0.128) & 0.878 (0.125)   \\
                               &                        & \textbf{GRACE}                 & 1.0 (0.0)     & 1.0 (0.0)     & 1.0 (0.0)     & 1.0 (0.0)     & 1.0 (0.0)     & 1.0   (0.0)     \\
                               &                        & \textbf{DeMem}                 & 0.796 (0.096) & 0.804 (0.105) & 0.82 (0.119)  & 0.835 (0.121) & 0.779 (0.124) & 0.783 (0.124)   \\ \cline{3-9}
                               & \multirow{4}{*}{\rotatebox{90}{URL}}   & \textbf{PME}                   & 0.928 (0.101) & 0.931 (0.103) & 0.912 (0.123) & 0.931 (0.095) & 0.872 (0.134) & 0.879   (0.132) \\
                               &                        & \textbf{MEMIT}                 & 0.89 (0.116)  & 0.894 (0.117) & 0.907 (0.11)  & 0.922 (0.094) & 0.833 (0.116) & 0.84 (0.12)     \\
                               &                        & \textbf{GRACE}                 & 1.0 (0.0)     & 1.0 (0.0)     & 1.0 (0.0)     & 1.0 (0.0)     & 1.0 (0.0)     & 1.0   (0.0)     \\
                               &                        & \textbf{DeMem}                 & 0.803 (0.101) & 0.811 (0.109) & 0.837 (0.121) & 0.862 (0.119) & 0.788 (0.11)  & 0.797 (0.113)   \\ \hline
\multirow{12}{*}{\rotatebox{90}{GPT-J 6B}}     & \multirow{4}{*}{\rotatebox{90}{email}} & \textbf{PME}                   & 0.945 (0.093) & 0.947 (0.096) & 0.954 (0.094) & 0.959 (0.09)  & 0.946 (0.096) & 0.95   (0.095)  \\
                               &                        & \textbf{MEMIT}                 & 0.902 (0.108) & 0.91 (0.107)  & 0.906 (0.124) & 0.916 (0.117) & 0.912 (0.118) & 0.914 (0.112)   \\
                               &                        & \textbf{GRACE}                 & 0.988 (0.06)  & 0.988 (0.059) & 1.0 (0.0)     & 1.0 (0.0)     & 0.997 (0.032) & 0.997   (0.029) \\
                               &                        & \textbf{DeMem}                 & 0.742 (0.06)  & 0.746 (0.077) & 0.749 (0.118) & 0.763 (0.121) & 0.726 (0.089) & 0.732 (0.096)   \\ \cline{3-9}
                               & \multirow{4}{*}{\rotatebox{90}{phone}} & \textbf{PME}                   & 0.953 (0.092) & 0.955 (0.09)  & 0.962 (0.082) & 0.966 (0.081) & 0.951 (0.096) & 0.956   (0.088) \\
                               &                        & \textbf{MEMIT}                 & 0.858 (0.116) & 0.864 (0.119) & 0.869 (0.136) & 0.883 (0.126) & 0.849 (0.121) & 0.859 (0.117)   \\
                               &                        & \textbf{GRACE}                 & 0.988 (0.06)  & 0.988 (0.059) & 1.0 (0.0)     & 1.0 (0.0)     & 0.997 (0.032) & 0.997   (0.029) \\
                               &                        & \textbf{DeMem}                 & 0.725 (0.041) & 0.732 (0.059) & 0.73 (0.112)  & 0.747 (0.11)  & 0.706 (0.094) & 0.722 (0.091)   \\ \cline{3-9}
                               & \multirow{4}{*}{\rotatebox{90}{URL}}   & \textbf{PME}                   & 0.935 (0.093) & 0.939 (0.093) & 0.904 (0.123) & 0.917 (0.111) & 0.898 (0.125) & 0.907   (0.119) \\
                               &                        & \textbf{MEMIT}                 & 0.853 (0.112) & 0.856 (0.115) & 0.878 (0.127) & 0.895 (0.114) & 0.833 (0.122) & 0.84 (0.124)    \\
                               &                        & \textbf{GRACE}                 & 0.988 (0.06)  & 0.988 (0.059) & 1.0 (0.0)     & 1.0 (0.0)     & 0.997 (0.032) & 0.997   (0.029) \\
                               &                        & \textbf{DeMem}                 & 0.723 (0.055) & 0.735 (0.071) & 0.734 (0.117) & 0.757 (0.123) & 0.694 (0.089) & 0.712 (0.091)   \\ \hline
\end{tabular}
\caption{Reliability of post-edit LLMs for all the considered baselines.}
\label{tab:post_edit_sim_all}
\end{table*}

\end{document}